\documentclass[twocolumn,hyperpdf,amsmath,amssymb,aps,prd,10pt,superscriptaddress,nofootinbib,noeprint,preprintnumbers,floatfix]{revtex4-2}

\usepackage[caption=false]{subfig}
\usepackage{graphicx, color}
\usepackage[letterspace=-10]{microtype} 

\usepackage{bm, amsmath, amsfonts}
\usepackage{multirow, tabularx, dcolumn,sidecap}

\usepackage[utf8]{inputenc} 
\usepackage{hyperref}
\usepackage{gensymb}
\usepackage[usenames,dvipsnames,svgnames,table]{xcolor} 
\usepackage{soul,color}
\usepackage{enumitem}
\usepackage{adjustbox}

\definecolor{jlab_red}{RGB}{192,39,45}
\definecolor{jlab_orange}{RGB}{249,102,0}
\definecolor{jlab_blue}{RGB}{47,122,121}
\definecolor{jlab_green}{RGB}{65,125,10}

\hypersetup{%
pdfsubject = {QCD},
pdfkeywords = {QCD, Hadron, Physics, Lattice, Meson, Scattering},
pdfauthor = {Hadron Spectrum Collaboration},
colorlinks = {true},
filecolor = {black},
linkcolor = {jlab_blue},
menucolor = {black},
citecolor = {jlab_green},
urlcolor = {jlab_green},
}{}

\usepackage[capitalize]{cleveref}

\usepackage{mfirstuc} %

\newcommand{\addReviewer}[2]{
  \expandafter\newcommand\csname #1\endcsname[1]{{\textbf{ \color{#2} \capitalisewords{#1}:\,##1}}}
  \expandafter\newcommand\csname #1cor\endcsname[2]{{\color{#2} \capitalisewords{#1}:\,\st{##1}{\, \textbf{##2}}}}
  \expandafter\newcommand\csname #1color\endcsname{#2}
  \expandafter\newcommand\csname #1todo\endcsname[1]{{\todo[inline,color=white!70!#2, caption={}]{\textbf{\capitalisewords{#1}}: ##1}}}
}

\usepackage{soul,color}
\definecolor{asparagus}{rgb}{0.53, 0.66, 0.42}
\usepackage{notoccite}

\addReviewer{arkaitz}{asparagus}

\makeatletter
\renewcommand\NAT@sort@cites[1]{%
  \let\NAT@cite@list\@empty
  \@for\@citeb:=#1\do{\expandafter\NAT@star@cite\@citeb\@@}%
  \if@filesw
    \expandafter\write\expandafter\@auxout
      \expandafter{\expandafter\string\expandafter\citation\expandafter{\NAT@cite@list}}%
  \fi
  \@ifnum{\NAT@sort>\z@}{%
    \expandafter\NAT@sort@cites@\expandafter{\NAT@cite@list}%
  }{}%
}%
\makeatother
\begin{document}

\preprint{JLAB-THY-23-3791}

\title{Determination of crossing-symmetric $\pi\pi$ scattering amplitudes and the quark mass evolution of the $\sigma$ constrained by lattice QCD}

\author{Arkaitz~Rodas}
\email{arodas@jlab.org}
\affiliation{\lsstyle Thomas Jefferson National Accelerator Facility, 12000 Jefferson Avenue, Newport News, VA 23606, USA}
\author{Jozef~J.~Dudek}
\email{dudek@jlab.org}
\affiliation{\lsstyle Thomas Jefferson National Accelerator Facility, 12000 Jefferson Avenue, Newport News, VA 23606, USA}
\affiliation{Department of Physics, College of William and Mary, Williamsburg, VA 23187, USA}
\author{Robert~G.~Edwards}
\email{edwards@jlab.org}
\affiliation{\lsstyle Thomas Jefferson National Accelerator Facility, 12000 Jefferson Avenue, Newport News, VA 23606, USA}

\collaboration{for the Hadron Spectrum Collaboration}
\date{\today}

\begin{abstract}
Lattice QCD spectra can be used to constrain partial-wave scattering amplitudes that, while satisfying unitarity, do not have to respect crossing symmetry and analyticity. This becomes a particular problem when extrapolated far from real energies, e.g. in the case of broad resonances like the $\sigma$, leading to large systematic uncertainties in the pole position. In this manuscript, we will show how dispersion relations can implement the additional constraints, using  as input lattice--determined $\pi\pi$ partial-wave scattering amplitudes with isospin--0,1,2. We will show that only certain combinations of amplitude parameterizations satisfy all constraints, and that when we restrict to these, the $\sigma$ pole position is determined with minimal systematic uncertainty. The evolution of the now well-constrained $\sigma$ pole with varying light quark mass is presented, showing how it transitions from a bound-state to a broad resonance.
\end{abstract}

\maketitle

\section{Introduction}
\label{sec:introduction}

Pion-pion scattering with isospin--0 plays a key role in nuclear and particle physics, with the $J^P=0^+$ component proving to be a feature of many low-energy phenomena of contemporary interest such as spontaneous symmetry breaking~\cite{Gell-Mann:1960mvl}, the long-range nucleon-nucleon interaction~\cite{Johnson:1955zz}, and hadronic contributions to the anomalous magnetic moment of the muon~\cite{Aoyama:2020ynm}. At low energies, this partial wave is dominated by the existence of the $\sigma$ particle, the lightest resonance in Quantum Chromodynamics (QCD). This extremely short-lived state which manifests as a slow energy variation of the scattering amplitude, has only had its complex energy plane pole position pinned down precisely relatively recently~\cite{Caprini:2005zr,Garcia-Martin:2011nna,Moussallam:2011zg}. How this state arises within QCD, and its relationship to other light-scalar mesons, like the $\kappa$, the $f_0(980)$, and the $a_0(980)$, remains unclear.

Our leading tool to study non-perturbative aspects of QCD, including hadron scattering and resonances, is Lattice QCD. This is a first-principles numerical approach in which only controlled approximations are made, and which features the ability to vary the value of quark masses and to explore the evolution of physical observables with such variation. Access to scattering amplitudes comes via computation of the discrete spectrum of states in the finite-volume of the lattice, extracted from the time-dependence of Euclidean correlation functions. A finite-volume quantization condition often referred to as the L\"uscher equation, relates the spectrum to infinite-volume partial-wave scattering amplitudes~\cite{Luscher:1986pf, Luscher:1990ux, Luscher:1991cf, Rummukainen:1995vs, He:2005ey, Christ:2005gi, Kim:2005gf, Guo:2012hv, Hansen:2012tf, Briceno:2012yi, Briceno:2014oea, Briceno:2017max}. In practice, the computed spectra are used to constrain unitarity-satisfying parameterizations of the isospin--0 \mbox{$S$--wave} partial-wave amplitude, which when analytically continued in the complex energy plane, yield the $\sigma$ pole~\cite{Briceno:2016mjc, Guo:2018zss, Mai:2019pqr}.

With light quark masses chosen so that the pion has a mass of 391 MeV, the $\sigma$ appears as a well-determined stable \emph{bound-state} pole below the $\pi\pi$ threshold. On the other hand, in calculations at a pion mass of 239 MeV it was found that a wide range of parameterizations were capable of describing the finite-volume spectrum, all giving compatible amplitude determinations across the elastic scattering region, but these various amplitudes feature resonant $\sigma$ pole positions lying deep in the complex plane which are scattered well outside the level of statistical uncertainty~\cite{Briceno:2016mjc}.

This observation is closely related to the longstanding challenge of accurately determining the $\sigma$ pole position from fits to experimental $\pi \pi$ elastic scattering data~\cite{Pelaez:2015qba}. In that case, a solution was found using \emph{dispersion relations} which implement the additional fundamental constraint of \emph{crossing symmetry}, leading to precise contemporary estimates of the $\sigma$ pole position~\cite{Caprini:2005zr,Garcia-Martin:2011nna,Moussallam:2011zg,ParticleDataGroup:2022pth}. 
Until recently there had only been a single attempt to apply crossing symmetry to $\pi \pi$ scattering amplitudes determined using lattice QCD~\cite{Guo:2018zss}. In that work the cross-channel effects were treated only perturbatively in a unitarized chiral perturbation theory framework.

In this manuscript we will apply dispersion relations, which allow us to implement crossing symmetry non-perturbatively, to lattice QCD data, considering two unphysical values of the light quark mass lying in the region where we believe the $\sigma$ is transitioning from being bound into being a broad resonance, showing that the dominant systematic error on the $\sigma$ pole position associated with parameterization of partial-wave scattering amplitudes can be almost entirely removed. Combining these results with calculations at heavier quark masses, we present a robust estimation within first-principles QCD of the evolution of the $\sigma$ pole with varying quark mass.

\vspace{3mm}

The Hadron Spectrum Collaboration ({\it hadspec}) has performed lattice calculations with two flavors of dynamical light quarks and a heavier dynamical strange quark ($N_f=2+1$) yielding discrete finite-volume spectra in channels with the quantum numbers of $\pi \pi$ scattering for four values of the light quark mass, corresponding to pion masses of $391, 330, 283$ and $239$ MeV~\cite{Dudek:2010ew,Dudek:2012gj,Dudek:2012xn,Wilson:2015dqa,Briceno:2016mjc,Briceno:2017qmb,Rodas:2023gma}. The $\pi\pi$ elastic partial-wave scattering amplitudes constrained using these spectra feature a $\rho$ in the isovector $P$--wave that is a resonance at all quark masses, while the isoscalar $S$--wave has an accurate interpretation only for the heaviest two quark masses, where the $\sigma$ appears as a well-determined bound-state pole. 
Other lattice calculations exist for $I=0$, with $N_f=2$~\cite{Guo:2018zss,Mai:2019pqr}, for $I=2$ with $N_f=2$~\cite{Culver:2019qtx}, and $N_f=2+1$~\cite{NPLQCD:2011htk,Bulava:2016mks,Akahoshi:2019klc}. The $\rho$ has been studied extensively both for $N_f=2$~\cite{CP-PACS:2007wro,Feng:2010es,Lang:2011mn,Pelissier:2012pi,Bali:2015gji,Guo:2016zos,Erben:2019nmx,Fischer:2020yvw}, $N_f=2+1$~\cite{CS:2011vqf,Sun:2015enu,Bulava:2016mks,Fu:2016itp,Alexandrou:2017mpi,Andersen:2018mau,Akahoshi:2021sxc} and for $N_f=2+1+1$~\cite{ExtendedTwistedMass:2019omo}. For a review on the topic, we refer the reader to Ref.~\cite{Mai:2022eur}.

In this manuscript, we will focus on the lightest two quark masses calculated by {\it hadspec}. In both these cases, the $I=2$ $S$--wave amplitude ($S2$) is weak, repulsive, and acceptably described over the elastic region by a scattering length approximation. The $I=2$ $D$--wave ($D2$) is extremely weak across the elastic region. The $I=1$ $P$--wave ($P1$) is dominated by the narrow $\rho$ resonance, which is well described by a Breit-Wigner amplitude, while the $F$--wave is completely negligible. The $I=0$ $S$--wave amplitude ($S0$) at both quark masses is a slowly varying function of energy, but there is a significant change in the behavior at threshold between the two quark masses, indicating a rapid variation in the scattering length with the quark mass. The generic functional forms used to describe these amplitudes can be found in~\cref{app:func_forms}, and in the original works mentioned above. Additionally, the $I=2$ channel at $m_\pi\sim 239$ MeV, which had not previously appeared in the literature, has been computed for this work and is presented in~\cref{app:860I2}.

The $S0$ `lattice data' (at both quark masses) proves to be describable by a large variety of unitarity-respecting parameterization forms, and these various descriptions lead to a spread of $\sigma$ pole estimates, with a scatter much larger than their statistical uncertainties, as seen in Ref.~\cite{Briceno:2016mjc}. It is clear, then, that the lattice data in $S0$ alone, even when it has rather small statistical errors, does not provide sufficient constraint to uniquely determine the $\sigma$ pole position. We will use the additional constraint of \emph{crossing symmetry}, in the form of dispersion relations applied to the coupled system of lattice data in all the partial waves above, to pin down the location of the $\sigma$ pole at $m_\pi \sim 283$ and $239$ MeV.

The manuscript is organized as follows: we briefly summarize the specific dispersion relations used in this work, applied to partial-wave amplitudes in~\cref{sec:DR}. In~\cref{sec:analysis}, we first define the metrics that quantify the degree to which our dispersion relations are fulfilled. Then, we apply these metrics to obtain a constrained system of amplitudes that respect all $S$-matrix principles. Once this set is obtained, in~\cref{sec:application}, we apply the dispersive outputs to the study of the subthreshold and resonance regions in~\cref{sec:subthreshold,sec:sigma_poles}, respectively, including discussions of the quark mass behavior of Adler zeroes and the $\sigma$ pole trajectory. Finally, we summarize our work in~\cref{sec:summary}.

\section{Dispersion relations} 
\label{sec:DR}

Generically, crossing symmetry relates two-particle scattering behavior in the $s,t$ and $u$--channels in terms of the same amplitude, $T(s,t,u)$, evaluated in different kinematical regions. For the case of $\pi\pi$ scattering with its three isospin configurations, given $T(s,t,u)$ we can construct the amplitudes of the different $s$-channel isospin processes as
\begin{align}
& T^{I=0}(s, t, u) = 3 T(s, t, u) + T(t, s, u) + T(u, t, s) ,\nonumber\\
& T^{I=1}(s, t, u) = T(t, s, u) - T(u, t, s) ,\nonumber\\
& T^{I=2}(s, t, u) = T(t, s, u) + T(u, t, s) \, .
\end{align}
All isospin amplitudes in $\pi \pi$ scattering are then related through simple algebraic crossing relations. However, crossing symmetry is obscured when amplitudes are partial-wave projected in one channel,
\begin{equation}
    t^I_\ell(s) = \tfrac{1}{64\pi} \int^1_{-1} d\!\cos \theta_s\, T^I(s,t,u)\, P_\ell( \cos \theta_s ),
\end{equation}
with the scattering dynamics in the $t$ and $u$--channels appearing as a \emph{left-hand cut}, i.e. as an imaginary part of $t^I_\ell(s)$ for $s < 0$. When the only constraint on the partial-wave amplitude is data for $s>s_\mathrm{thr}$, typically the left-hand cut remains undetermined.

In the case that a partial-wave is dominated by a \emph{narrow} resonance (like the $\rho$), the impact of the left-hand cut singularity for real energies above threshold is typically negligible, since the nearby resonance pole dominates. However, when there is no resonance, or when there is a broad resonance, deep in the complex plane (like the physical $\sigma$), the left-hand cut contribution can be significant, and there is a need to constrain it accurately. 

\begin{figure*}[!hbt]
\includegraphics[width=.99\textwidth]{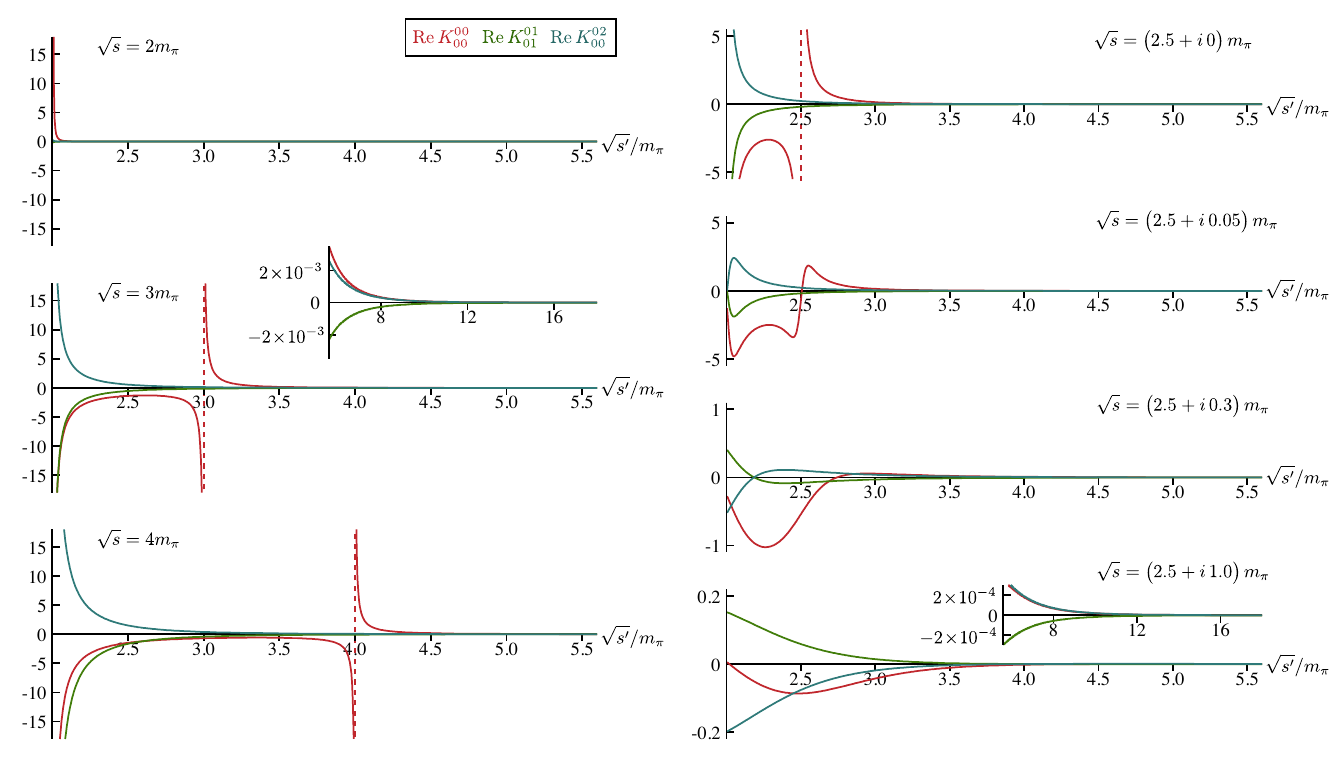}
\caption{
Real part of kernel functions in \cref{eq:DR} contributing to dispersion of $S0$, $\mathrm{Re}\, K^{0I'}_{0\ell'}$, plotted in units where $1$ on the $y$-axis represents a value $4 m_\pi^2$. Left column: functions evaluated on the real $s'$--axis for three values of $s$ -- inset indicates the behavior in the high-energy region. Right column: Kernel functions evaluated for complex values of $s$.
}\label{fig:RoyKernels}
\end{figure*}

In the case of $\pi\pi$ elastic scattering, the left-hand cuts of the partial-waves with $I=0,1,2$, $t^I_\ell(s)$, can be constrained by crossing symmetry, and \emph{dispersion relations} provide an approach to implement this symmetry using the analytic properties of the amplitudes $T^I(s,t,u)$. Such relations are obtained by application of Cauchy's theorem to $T^I(s,t,u)$~\cite{Martin:1970}, introducing subtractions to ensure that the integrals converge at infinity. Upon partial-wave projection, the dispersion relations take the form,
\begin{equation}
\tilde{t}^I_\ell(s) = \tau^I_\ell(s) 
+ \sum_{I', \ell'} 
 \int_{4m_\pi^2}^\infty \!\!\! ds' \, K^{II'}_{\ell\ell'}\!(s'\!,s)\; \mathrm{Im}\, t^{I'}_{\ell'}(s') \, ,
\label{eq:DR}
\end{equation}
where scattering amplitudes consistent with crossing symmetry and unitarity will have $\tilde{t}^I_\ell(s) = t^I_\ell(s)$. The $\{ \tau^I_\ell(s) \}$ are low-order polynomials in $s$ featuring a number of free parameters set by the number of subtractions, while the kernels $K^{II'}_{\ell\ell'}\!(s'\!,s)$ are known functions (also depending upon the number of subtractions, see Refs.~\cite{Ananthanarayan:2000ht,Garcia-Martin:2011iqs}) that encode the crossing relations\footnote{The diagonal kernels, $K^{II}_{\ell\ell}(s',s)$, contain a simple pole $\sim \frac{1}{\pi} \frac{1}{s'-s}$ which ensures that $\mathrm{Im}\, \tilde{t}^I_\ell(s)$ is always exactly equal to $\mathrm{Im}\, t^{I}_{\ell}(s)$ for real energies above threshold.}. The structure of the kernels ensures that the dispersed amplitudes have left-hand cuts set by the dynamics in the crossed-channels.

For \emph{twice-subtracted} dispersion relations, commonly referred to as ``Roy'' equations~\cite{Roy:1971tc}, the functions $\tau^I_\ell(s)$ take the form
\begin{align}
 \tau^0_0(s) / m_\pi &= \tfrac{1}{3} \big( a^0_0 + 5 a^2_0  \big) + \tfrac{1}{3} \big( 2 a^0_0 - 5 a^2_0  \big) \frac{s}{4m_\pi^2} \, ,   \nonumber \\
 \tau^1_1(s) / m_\pi &= \tfrac{1}{18} \big( 2a_0^0 - 5 a^2_0 \big) \frac{s - 4 m_\pi^2}{4 m_\pi^2} \, ,\nonumber \\
 \tau^2_0(s) / m_\pi &= \tfrac{1}{6} \big( 2 a^0_0 +  a^2_0  \big) - \tfrac{1}{6} \big( 2 a^0_0 - 5 a^2_0  \big) \frac{s}{4m_\pi^2} \, , 
  \label{eq:tau_roy}
\end{align}
where the $S$-wave scattering-lengths, 
\begin{equation}
    a^I_0 \equiv \mathrm{Re}\, t^I_0(s=4m_\pi^2) / m_\pi,
\end{equation}
are parameters that can be fixed using the $S0$, $S2$ amplitude behaviors at threshold, since the contribution of the integrals in~\cref{eq:DR} goes to zero as $\sqrt{s}\to 2m_\pi$.

\vspace{2mm}
Figure~\ref{fig:RoyKernels} illustrates the dispersive kernel functions relevant for obtaining the $S0$ dispersed amplitude, $K^{0I'}_{0\ell'}\!(s'\!,s)$, where we observe  that crossed-channels will have an influence at elastic scattering energies, and that all contributions from amplitudes at high energies are heavily suppressed due to the subtraction scheme. The right column shows the kernels evaluated at \emph{complex} values of $s$ where we observe the pole singularity at real $s'=s$ being smoothed out. The lowest entry in the right column is evaluated in the region where the $\sigma$ pole is eventually found for $m_\pi \sim 239 \, \mathrm{MeV}$, and here we observe that the influence of the crossed-channels, $P1$ and $S2$ (which contribute to the \emph{left-hand cut}), will be significant.

The integrals in~\cref{eq:DR} run over all energies above the elastic scattering threshold, and inevitably our knowledge of $\{ t^I_\ell(s) \}$ stops at some point, so it proves necessary to parameterize the high-energy behavior, which we may do using known Regge asymptotics. 
In our implementation, we divide the integration from threshold to infinity into two parts: an integration from threshold to $s_h$ where lattice constrained amplitudes are used, and an integration from $s_h$ to infinity where Regge parameterizations are used.
For both cases in our work, $m_\pi \sim 239, \,283$ MeV, we use $\sqrt{s_h} = 0.22 \, a_t^{-1}$, which for $m_\pi \sim 239$ MeV corresponds to $\sqrt{s_h} = 5.6\,  m_\pi = 1.34 \, \mathrm{GeV} $. Details of the high-energy Regge parameterizations which model $t$-channel exchanges in the amplitudes $T^I(s,t,u)$ are presented in~\cref{app:dr_input}.

The dispersed amplitudes $\tilde{t}^I_\ell(s)$ in~\cref{eq:DR} are obtained by evaluating the integrals numerically. The input lattice-constrained partial-wave amplitudes carry statistical uncertainties associated with being constrained by energy levels computed on a finite ensemble of lattice configurations. These are available in the form of (correlated) statistical uncertainties on the parameters of each amplitude parameterization.
In addition, as discussed in~\cref{app:regge}, the high-energy Regge parameterized amplitudes carry an assigned conservative uncertainty. Given a set of fits to lattice data with parameters $\{a_i\}$, and Regge contributions with a fixed fractional uncertainty, the dispersion relation uncertainties $\Delta \tilde{f}^I_\ell(s,\{a_i\})$ are obtained by linearized (correlated) error propagation at each sampled energy point.

In this work, we will report results using \emph{minimally-subtracted} relations (often referred to as ``GKPY equations''~\cite{Garcia-Martin:2011iqs}), and \emph{twice-subtracted} relations (``Roy equations''~\cite{Roy:1971tc}). A choice can be made to implement \emph{more} than the minimal number of subtractions needed for convergence, which will have the practical consequence of reducing the contribution of the partial-wave amplitudes at high-energies, in particular in the region where they are Regge-parameterized, in exchange for increased sensitivity to the behavior of the amplitudes at the subtraction point, which is typically the elastic threshold. We observe that the \emph{twice-subtracted} amplitudes are minimally sensitive to how we parameterize the high-energy behavior, and while the \emph{minimally-subtracted} amplitudes do show sensitivity to estimated high-energy behavior, nevertheless, compatible results for amplitudes at low energies are obtained. 
In the main body of the paper we will present only results applying \emph{twice-subtracted} dispersion relations, where, as shown in~\cref{eq:tau_roy}, the $\tau^I_\ell(s)$ are linear in $s$ and depend only upon the $S$--wave scattering lengths, $a_0^0$, $a_0^2$. Results using \emph{minimally-subtracted} dispersion relations are presented in~\cref{app:GKPY}.

\section{Dispersive evaluation of lattice constrained amplitudes}
\label{sec:analysis}

As described above, the discrete energy spectra extracted from lattice QCD calculations constrain definite isospin partial-wave amplitudes one-by-one in the elastic scattering region in a way that exactly respects unitarity, but which has no constraint from crossing symmetry. Typically multiple parameterizations prove to be capable of describing the lattice data for real energies above threshold. We will establish which combinations of partial-wave amplitude parameterizations are compatible with first-principles using dispersion relations. Upon input of a selected set of lattice amplitudes, $\big\{ t_\ell^I(s) \big\}$, into the right-hand-side of~\cref{eq:DR}, a set of \emph{dispersed amplitudes}, $\big\{\tilde{t}_\ell^I(s) \big\}$, are produced which have the same imaginary parts, but modified real parts. In order for these amplitudes to be compatible with unitarity, they must be statistically compatible with the input amplitudes, $\big\{ t_\ell^I(s) \big\}$, in the elastic scattering region.

We assess the suitability of the dispersed amplitudes using two metrics. The first, which we call $d^2$, compares the real part of the dispersed amplitude in one partial-wave ($\tilde{f}^I_\ell(s) \equiv \mathrm{Re}\, \tilde{t}_\ell^I(s)$) with the real part of the input lattice amplitude  (${f}^I_\ell(s) \equiv \mathrm{Re}\, {t}_\ell^I(s)$),
\begin{equation}
\big[ d^2 \big]^I_\ell \equiv \sum_{i=1}^{N_\mathrm{smpl}} 
\left( 
   \frac{\tilde{f}^I_\ell(s_i) - f^I_\ell(s_i) }
   {\Delta \big[ \tilde{f}^I_\ell(s_i) - f^I_\ell(s_i)  \big]} 
\right)^2 \, , \label{eq:dsq}
\end{equation}
where the difference is sampled at a large number of equally spaced energy values in the elastic scattering region, and where the uncertainty in the difference is computed by linearly propagating the correlated uncertainties on the lattice-fit amplitude parameters, and in addition, the conservative uncertainty placed on the high-energy Regge-like behavior (see~\cref{app:regge} for details). 
\begin{figure*}[!ht]
\includegraphics[width=.95\textwidth]{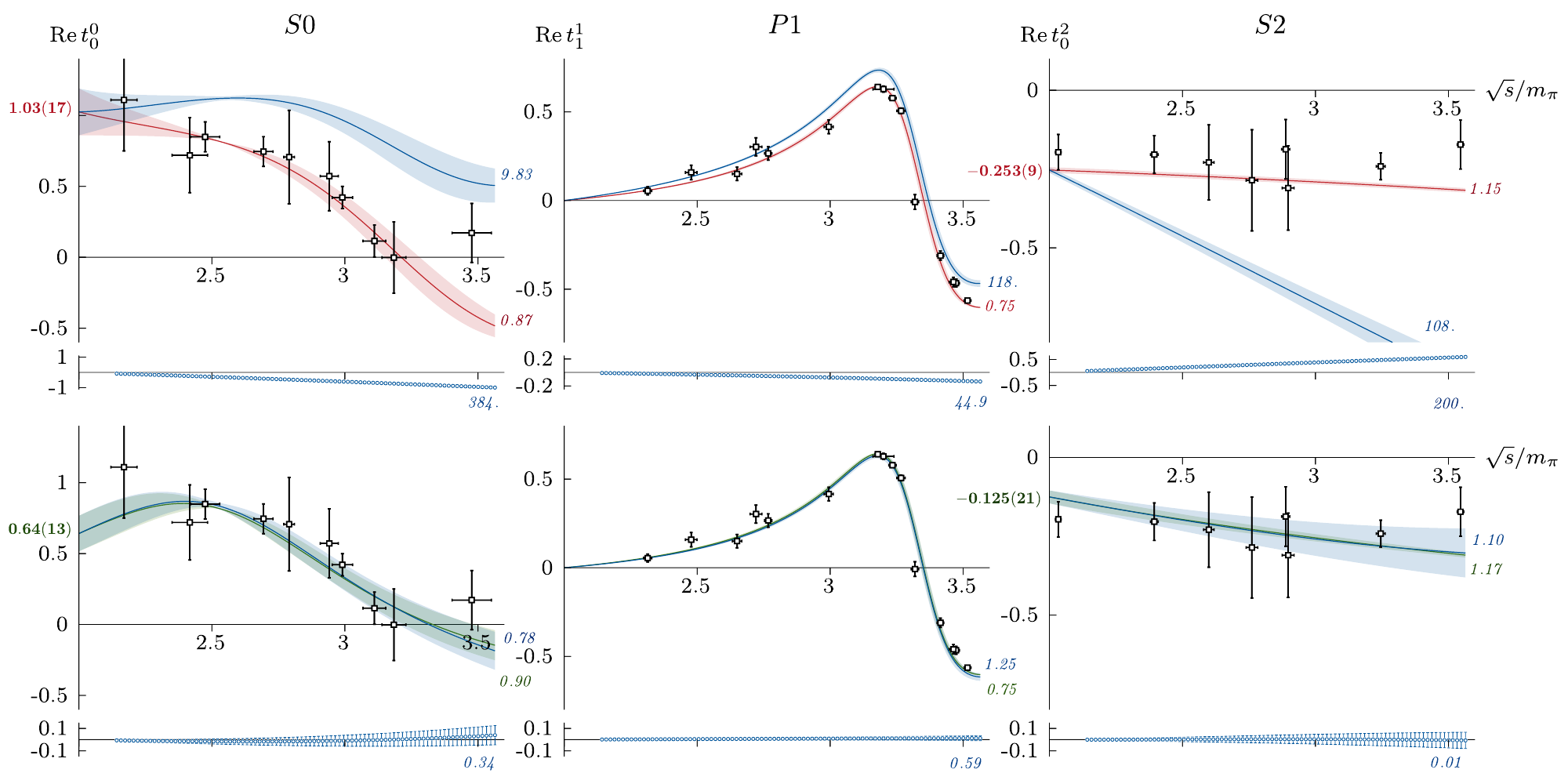}
\caption{Real parts of $S0$, $P1$, $S2$ scattering amplitudes for $m_\pi \sim 239$ MeV. Data-points indicate the constraint provided by discrete lattice QCD spectra. The top row show an example of parameterizations (red) which describe well the lattice data, but which generate dispersed amplitudes (blue) that are in poor agreement, leading to large values of $d^2$ and $\tilde{\chi}^2$. The second rows show a case where the lattice amplitudes (green) prove to be in good agreement with the dispersed amplitudes. The colored numbers with an error estimate show the $S$-wave scattering lengths in $m^{-1}_\pi$ units. The red(green) italic numbers show the $\tilde{\chi}^2/N_{\text{lat}}$ values for the input amplitudes, whereas the blue numbers on those same panels show the  $\tilde{\chi}^2/N_{\text{lat}}$ values for the dispersed amplitude.
Beneath each panel, values of the quantity in the large parentheses in Eq.~\ref{eq:dsq} at $N_\mathrm{smpl}= 91$ points are given, and the numbers listed there show the $d^2/N_{\text{smpl}}$ values.
}\label{fig:amp860}
\end{figure*}

\begin{figure*}[!ht]
\includegraphics[width=.95\textwidth]{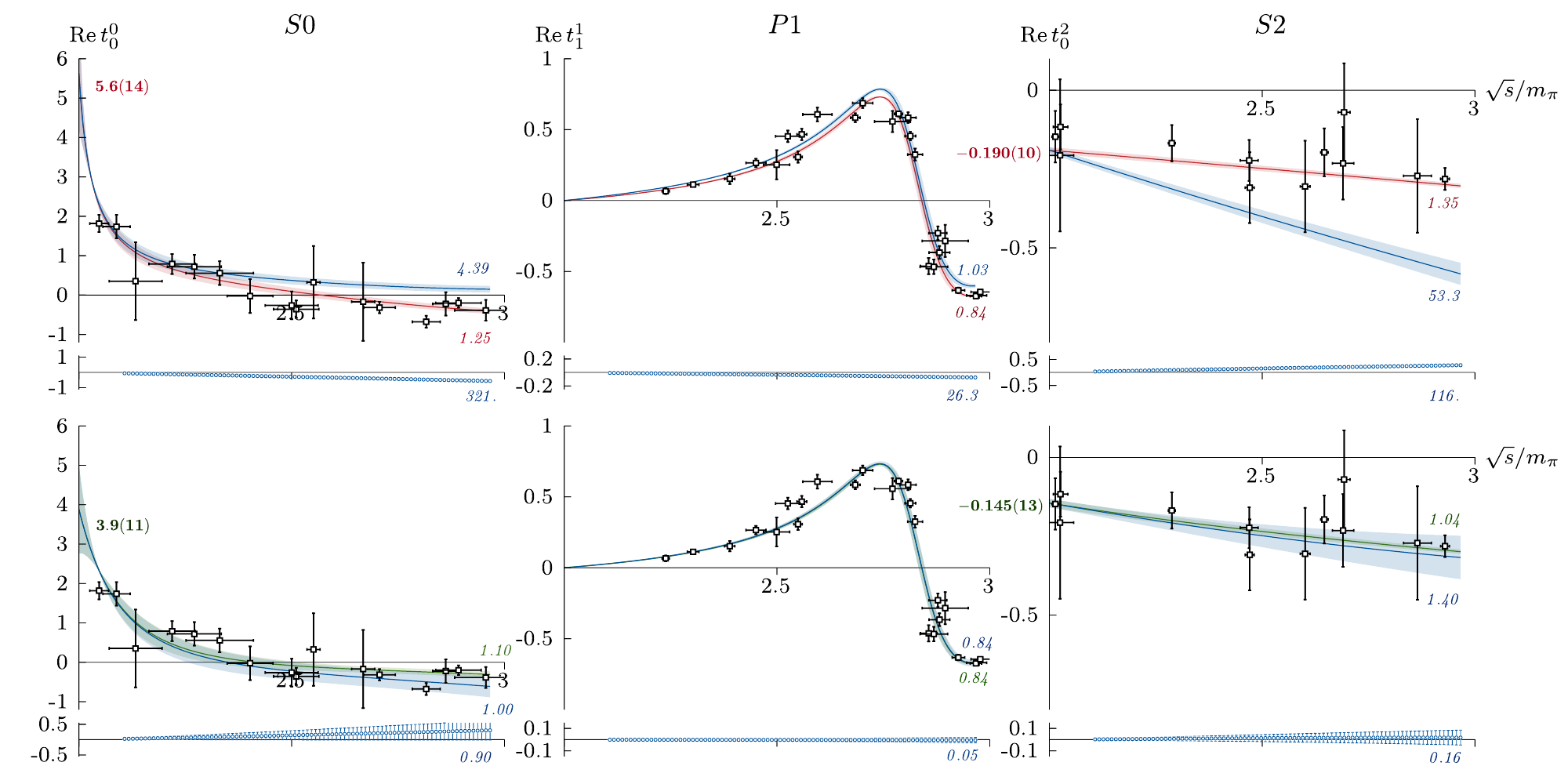}
\caption{As Figure~\ref{fig:amp860} but for $m_\pi \sim 283$ MeV. The lower panels sampled at $N_\mathrm{smpl}= 89$ points.}\label{fig:amp856}
\end{figure*}

The $d^2$ metric can yield small values when at least one of $f_\ell^I$ or $\tilde{f}_\ell^I$ has large statistical uncertainties, corresponding to an (undesired) imprecise amplitude description, so we choose to supplement it with a second $\chi^2$--like metric which compares the central value of the dispersed amplitude with the lattice data,
\begin{align}
\big[ \tilde{\chi}^2 \big]^I_\ell &\equiv \sum_{i,j=1}^{N_\mathrm{lat}} 
\left( \tfrac{ \mathfrak{f}_i - \tilde{f}^I_\ell(s_i)}{\Delta_i} \right)
\mathrm{corr}(\mathfrak{f}_i, \mathfrak{f}_j)^{-1}
\left( \tfrac{ \mathfrak{f}_j - \tilde{f}^I_\ell(s_j)}{\Delta_j} \right) \, , 
\label{chisq}
\end{align}
where $\mathfrak{f}_i$ are the discrete values of $\mathrm{Re}\, t^I_\ell$ extracted from solving the L\"uscher finite-volume condition at energy ${E_i = \sqrt{s_i}}$. The construction of the corresponding uncertainty, $\Delta_i$, is described in Appendix~\ref{app:extended_values}. 

\begin{figure*}[!htb]
\adjustbox{valign=c}{%
\begin{minipage}[t]{.7\textwidth}
  \centering
  \subfloat{\includegraphics[width=.99\textwidth]{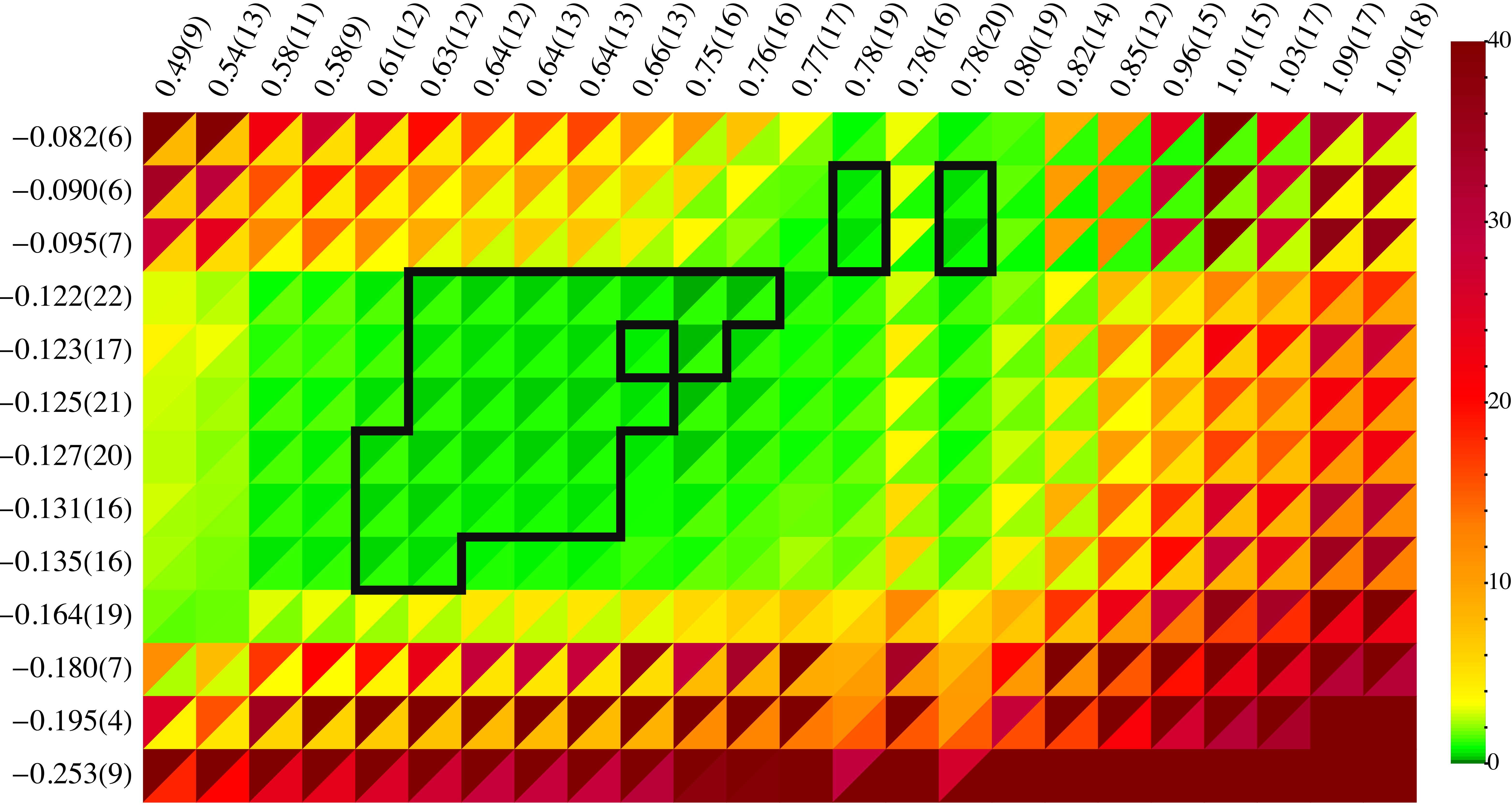}}
\end{minipage}}
\hspace{.5cm}\adjustbox{valign=c}{%
\begin{minipage}[t]{.25\textwidth}
\caption{\label{fig:roy_select860} For $m_\pi\sim 239$ MeV, for each combination of parameterizations of $S0$ (columns) and $S2$ (rows), each ordered by the magnitude of their scattering length in $m^{-1}_\pi$ units, boxes are colored according to the average value of $d^2/N_\mathrm{smpl}$ (upper triangle) and $\tilde{\chi}^2/N_\mathrm{lat}$ (lower triangle) over $S0$, $P1$, $S2$. The region indicated by the black outline shows parameterization combinations that have all amplitudes having metric values below the cutoffs presented in the text.}
\end{minipage}}%
\end{figure*}

\begin{figure*}[!htb]
\adjustbox{valign=c}{%
\begin{minipage}[t]{.7\textwidth}
  \centering
  \subfloat{\includegraphics[width=.99\textwidth]{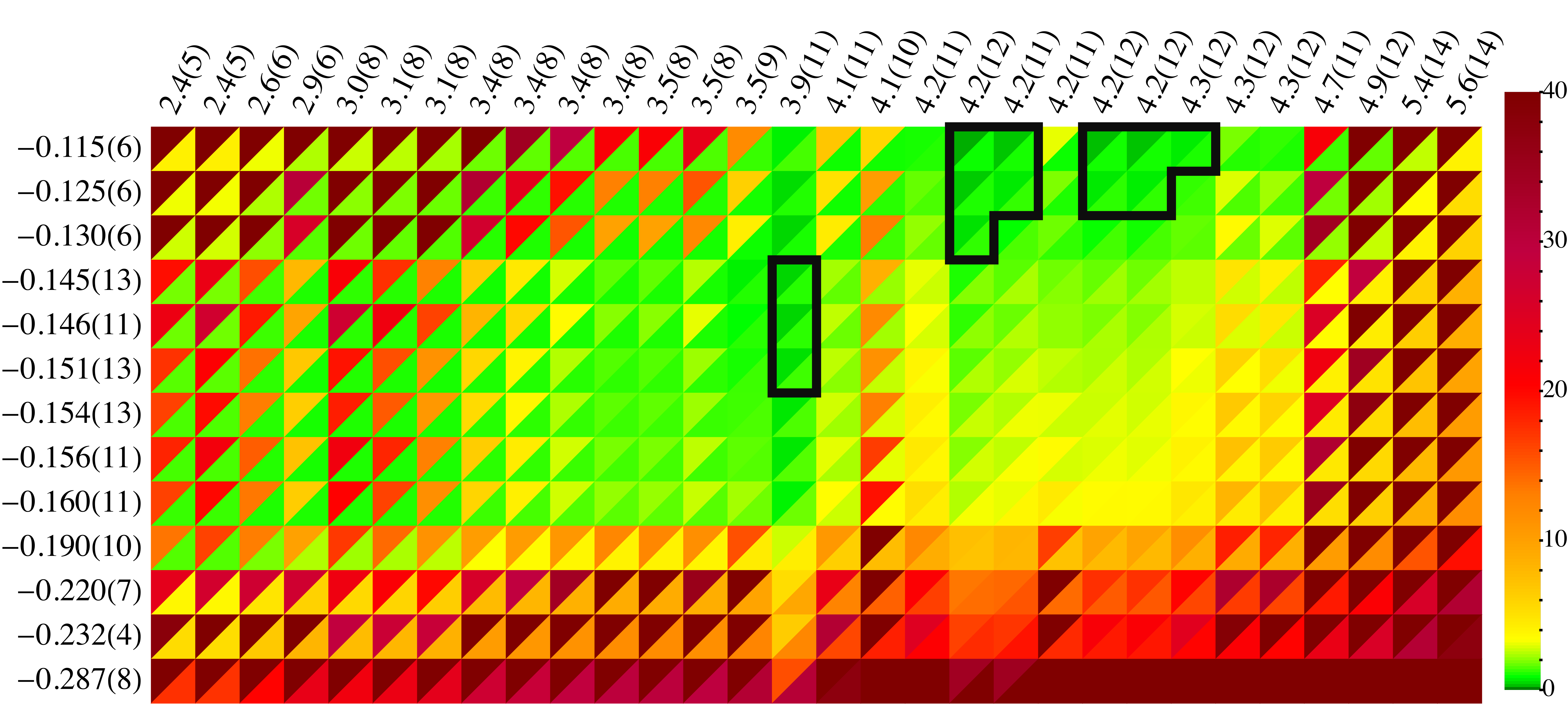}}
\end{minipage}}
\hspace{.5cm}\adjustbox{valign=c}{%
\begin{minipage}[t]{.25\textwidth}
\caption{\label{fig:roy_select856} As Figure~\ref{fig:roy_select860} but for $m_\pi\sim 283$ MeV.}
\end{minipage}}%
\end{figure*}

Figures~\ref{fig:amp860}, \ref{fig:amp856}, show examples of $m_\pi \sim 239$ MeV and $m_\pi \sim 283$ MeV $S0$, $P1$ and $S2$ lattice amplitudes~\footnote{see Refs.~\cite{Briceno:2016mjc,Rodas:2023gma} and~\cref{app:func_forms} for details of the particular functional forms used.}, and their dispersively modified counterparts, illustrating how the metrics defined above select consistent and reject inconsistent combinations. 
In practice, we find that there is little variation in the metric values with change of $P1$ parameterization, so we opt to fix this to one particular successful form for both analyses~\footnote{a $K$-matrix with a single pole plus a constant, and a dispersed phase-space subtracted so that its real part vanishes at the bare $K$-matrix pole.}. We compute dispersed amplitudes in the remaining space of choices of $S0$, $S2$ parameterizations, and retain only those combinations which have $d^2/N_\mathrm{smpl} < 1$ and $\tilde{\chi}^2/N_\mathrm{lat} < 2$ for all partial-waves, $S0, P1, S2$~\footnote{The sensitivity to this particular choice is investigated in~\cref{app:extended_values}.}. 

The values of the metrics for a large number of $S0$, $S2$ amplitude parameterizations are shown in \cref{fig:roy_select860}~(for $m_\pi \sim 239$ MeV) and \cref{fig:roy_select856}~(for $m_\pi \sim 283$ MeV), where it is clear that applying cuts on the metrics, and thus enforcing both unitarity and crossing symmetry, leads to a much reduced set of amplitudes relative to those which acceptably described the lattice energy levels in a conventional ``partial-wave--by--partial-wave'' analysis. The value of the relevant $S$--wave scattering length is provided for each parameterization, and it is clear that this approach has significantly reduced the range of acceptable values of $a_0^0$ and $a_0^2$. 

In the $m_\pi \sim 283$ MeV case shown in \cref{fig:roy_select856}, $a_0^0$ is large and positive, of much larger magnitude than in the $m_\pi \sim 239$ MeV case, while $a_0^2$ remains small. Equations~(\ref{eq:DR}) and (\ref{eq:tau_roy}) are such that $a_0^0$ and $t^0_0(s > 4 m_\pi^2)$ feature for \emph{all} partial-waves, and in order to get lineshapes compatible with data, including weak scattering  in $I=2$, a delicate cancellation between the integral over $t^0_0(s)$ and the contribution of $a^0_0$ is required. This is reflected in the relatively small number of amplitudes found with small metric values.

\section{Dispersed amplitudes away from the elastic scattering region}
\label{sec:application}

The acceptable dispersed amplitudes, $\tilde{t}^I_\ell(s)$, found in the previous section (by virtue of small values of metrics), have several desirable properties, owing to the fact that they effectively respect unitarity, analyticity \emph{and} crossing symmetry.

An illustration of this comes in their behavior for real energies below the elastic threshold, which for the lattice-constrained input amplitudes was essentially unconstrained and varied wildly between parameterizations, with some even featuring unphysical singular behavior. In a ``partial-wave--by--partial-wave'' analysis, this subthreshold region is obtained only as an extrapolation, and should not be expected to be accurate away from the data above or near the threshold. 
On the other hand, our dispersed amplitudes, $\{\tilde{t}^I_\ell(s)\}$, implicitly include information both from the right- and the left-hand cuts (via crossing), and we are in effect \emph{interpolating} between these, with the additional constraint of analyticity being imposed by the use of dispersion relations which originate in the Cauchy theorem.
Similarly the extrapolation into the complex energy plane, where resonance poles are expected to be found, should be rendered more stable.

\subsection{Subthreshold region}
\label{sec:subthreshold}

\begin{figure*}[!htb]
\raisebox{-0.5\height}{\includegraphics[width=0.45\textwidth]{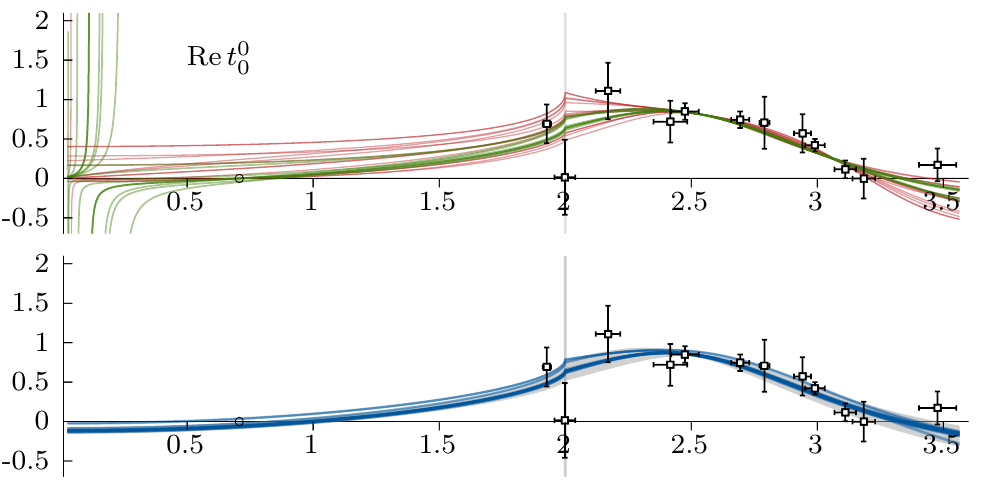}}
\hspace*{5mm}
\raisebox{-0.5\height}{\includegraphics[width=0.45\textwidth]{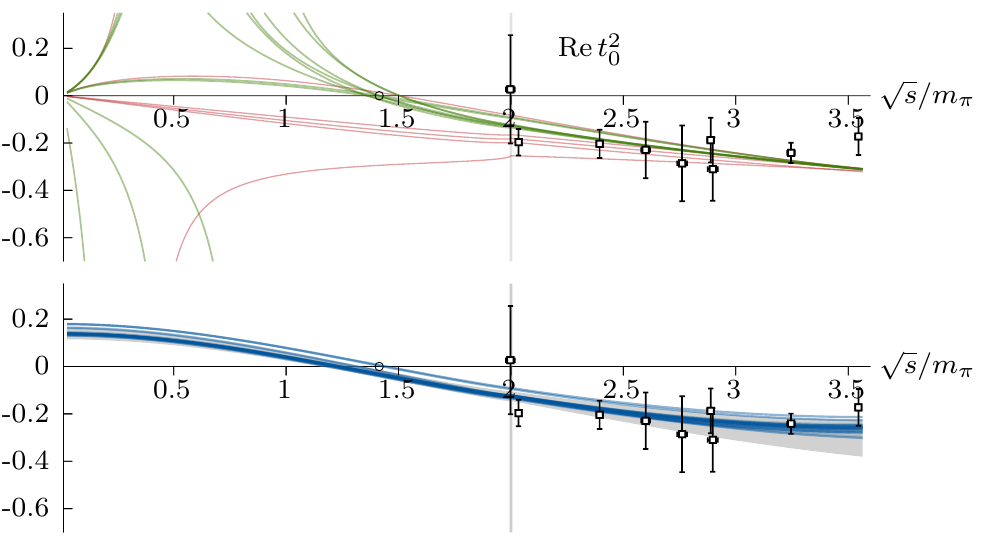}}
\caption{
$S$-wave amplitudes for $m_\pi \sim 239$ MeV. Top panel: Central values of the real parts of $S0$ and $S2$, both above and below the two-pion threshold, for all amplitude parameterizations. Those plotted in green correspond to input parameterizations that produce at least one combination respecting the metric cuts presented above, while all others are plotted in red. Open circles on axis indicate the locations of Adler zeroes in the leading-order of $\chi$PT. Bottom panel: Real parts of the corresponding dispersive amplitudes for all amplitude combinations respecting the metric cuts presented above. The uncertainty on one example amplitude is shown by the gray band.
}\label{fig:subthreshold860}
\end{figure*}

\begin{figure*}[!htb]
\raisebox{-0.5\height}{\includegraphics[width=0.45\textwidth]{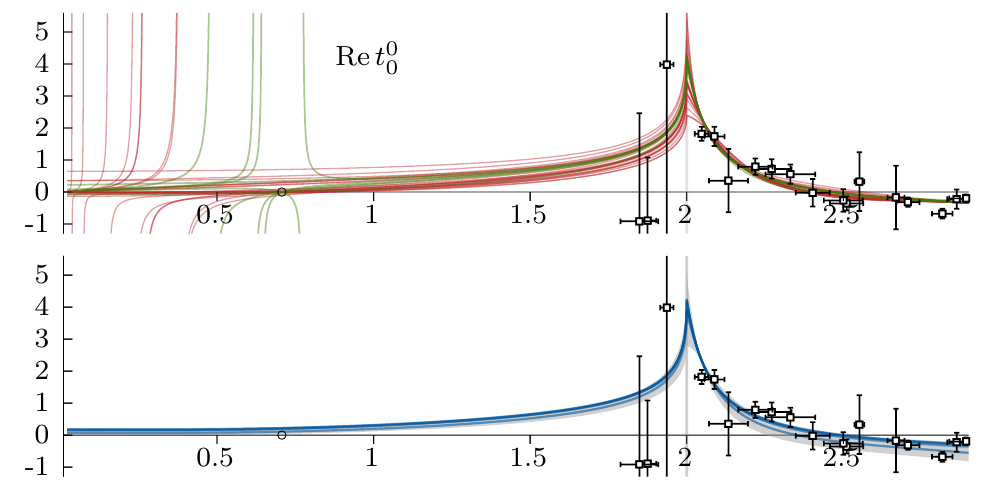}}
\hspace*{5mm}
\raisebox{-0.5\height}{\includegraphics[width=0.45\textwidth]{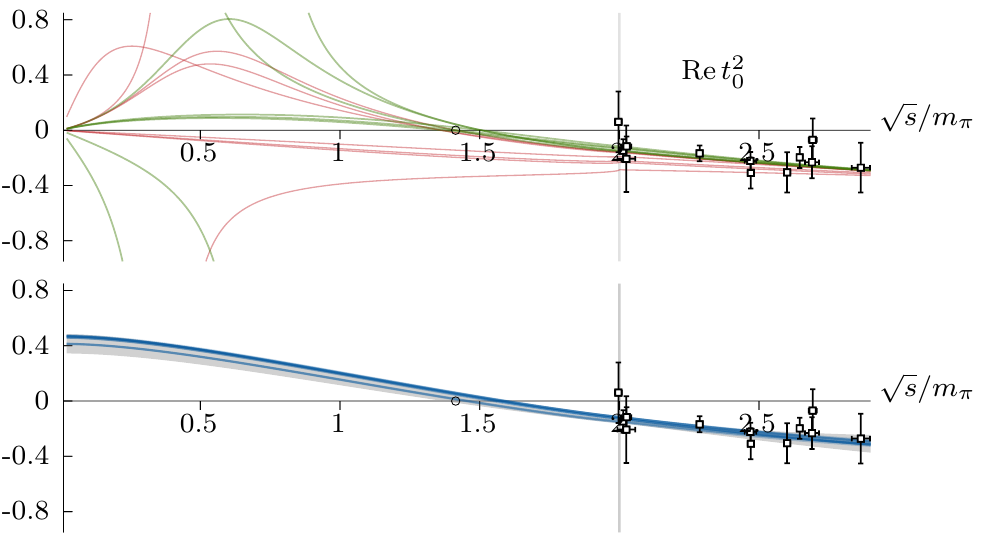}}
\caption{
As Figure~\ref{fig:subthreshold860} but for $m_\pi \sim 283$ MeV.
}\label{fig:subthreshold856}
\end{figure*}

The amplitude parameterizations used to describe lattice QCD spectra above threshold (and sometimes slightly below threshold) in a particular partial-wave are built to exactly obey unitarity above threshold, but they are not typically guaranteed to be free of unphysical behavior far below threshold. Usually this is excused as irrelevant since the amplitudes are only needed in a small region around the real energy axis above threshold (for example when narrow resonances are sought), but if extrapolation further into the complex energy plane is required, such parameterization artifacts may be problematic.

The use of dispersion relations can remedy this problem. Since the dispersion relations take as input the amplitudes {\it only} above threshold (where they are well constrained), while the subthreshold behavior is controlled by the kernel functions (which have correct analytic properties), the behavior of the dispersed amplitudes below threshold is rendered free of singularities.

This is illustrated in Figures~\ref{fig:subthreshold860}, \ref{fig:subthreshold856}, where the $S0$ and $S2$ amplitudes are shown for $m_\pi \sim 239,\,283$ MeV. The upper panels in each case shows the input lattice amplitudes (where amplitudes that systematically fail the metric cuts are shown in red) where subthreshold divergences are observed to be present, as is a significant scatter of behavior such that one can argue that the lattice data (above threshold, in a single partial-wave) has not constrained in any reliable way the amplitude behavior far below threshold. On the other hand, in the lower panels, we observe that all dispersed amplitudes satisfying the metric cuts show broadly compatible singularity-free behavior below threshold. The scatter of behaviors of acceptable amplitudes is observed to be at the level of the uncertainty (shown for one example amplitude by the gray band).

In \cref{fig:subthreshold860}, for $m_\pi \sim 239$ MeV, both sets of dispersed amplitudes are observed to feature a zero-crossing below threshold, located near to $s/m_\pi^2 \approx 0.8$ for $S0$ and $s/m_\pi^2 \approx 1.6$ for $S2$. The presence of such zeroes, known as ``Adler zeroes'', is an expectation of chiral perturbation theory~\cite{Weinberg:1978kz}, and we show in the figure the expected location at leading order ($s/m_\pi^2 = \tfrac{1}{2}$ for $S0$ and $s/m_\pi^2 = 2$ for $S2$). The zeroes in the dispersed amplitudes are observed to differ from these expectations, with a non-negligible spread.

In \cref{fig:subthreshold856}, for $m_\pi \sim 283$ MeV, the $S2$ amplitude is seen to feature an Adler zero near $s/m_\pi^2 \approx 2.3$, while the $S0$ amplitude does not appear to cross zero between the left-hand cut and threshold, in contradiction to the expectations of leading-order $\chi$PT. As such, we would argue that analyses of lattice QCD obtained spectra using amplitudes which  \emph{enforce} an Adler zero fixed at the leading order location are potentially introducing a systematic bias and this may impact results such as scattering lengths or low-lying pole positions.

\subsection{Resonance poles in the dispersed amplitudes}
\label{sec:sigma_poles}

\begin{figure*}[!hbt]
\resizebox{\textwidth}{!}{
   \raisebox{-0.5\height}{\includegraphics[width=.52\textwidth]{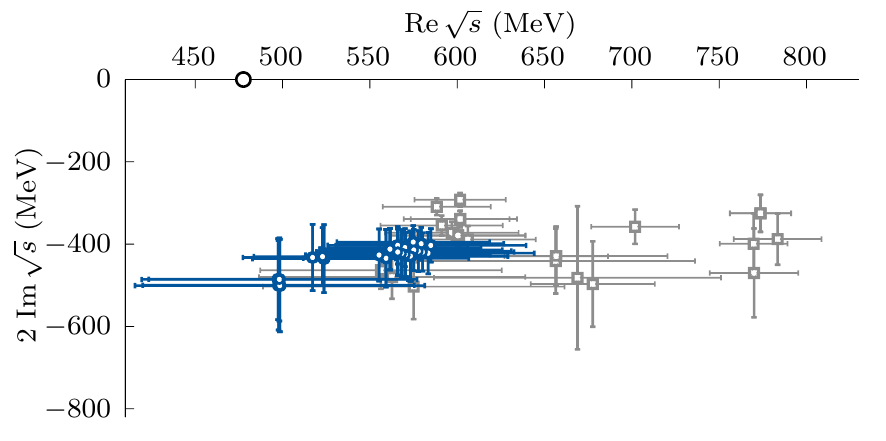}} \raisebox{-0.5\height}{\includegraphics[width=.45\textwidth]{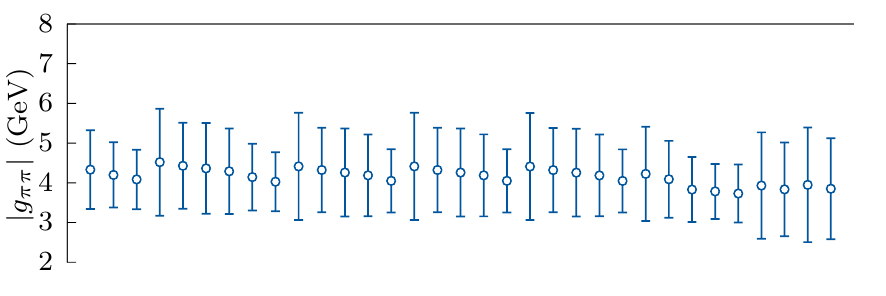}}
   }
 \caption{For $m_\pi \sim 239$ MeV, pole locations in $S0$ amplitude for lattice-fit amplitudes (gray) and dispersed amplitudes satisfying the metric cuts described in the text (blue), and modulus of the couplings (as defined in Refs.~\cite{Garcia-Martin:2011nna,Rodas:2023gma}) extracted from the pole residues.}
\label{fig:sigma_poles860}
\end{figure*}

\begin{figure*}[!hbt]
\resizebox{\textwidth}{!}{
   \raisebox{-0.5\height}{\includegraphics[width=.52\textwidth]{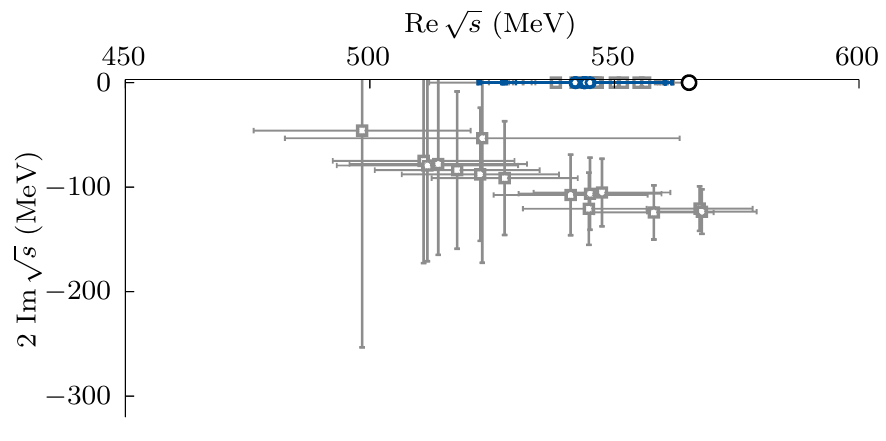}} \raisebox{-0.5\height}{\includegraphics[width=.45\textwidth]{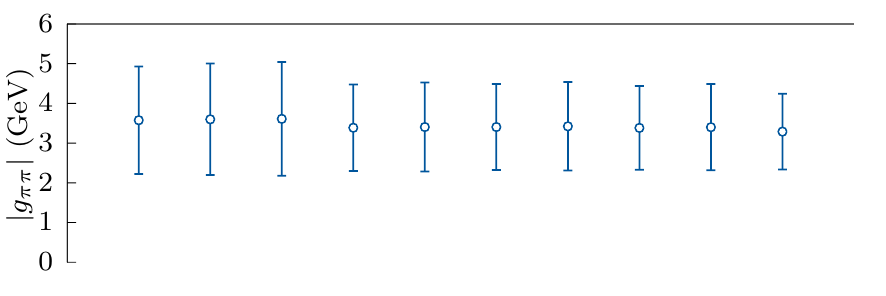}}
   }
\caption{As Figure~\ref{fig:sigma_poles860} but for $m_\pi \sim 283$ MeV. Cases in which the pole locations feature very large uncertainties have been excluded for clarity of presentation.}
\label{fig:sigma_poles856}
\end{figure*}

The location of the $\rho$ resonance pole in the dispersed $P1$ amplitudes is found to be compatible with the small spread observed in the input lattice amplitudes, as expected for a narrow resonance. On the other hand, for the $\sigma$ pole in $S0$, which at $m_\pi \sim 239$ MeV is lying deep in the complex plane, the acceptable dispersed amplitudes all have a pole that lies in a much-reduced region, as shown in \cref{fig:sigma_poles860}. As hoped, the imposition of analyticity and crossing symmetry, constrained by lattice data in all relevant isospins and low partial-waves, has led to a robust extraction of the $\sigma$ pole position, which is observed to be independent of any significant parameterization dependence.

\begin{figure*}[!hbt]
\includegraphics[width=.6\textwidth]{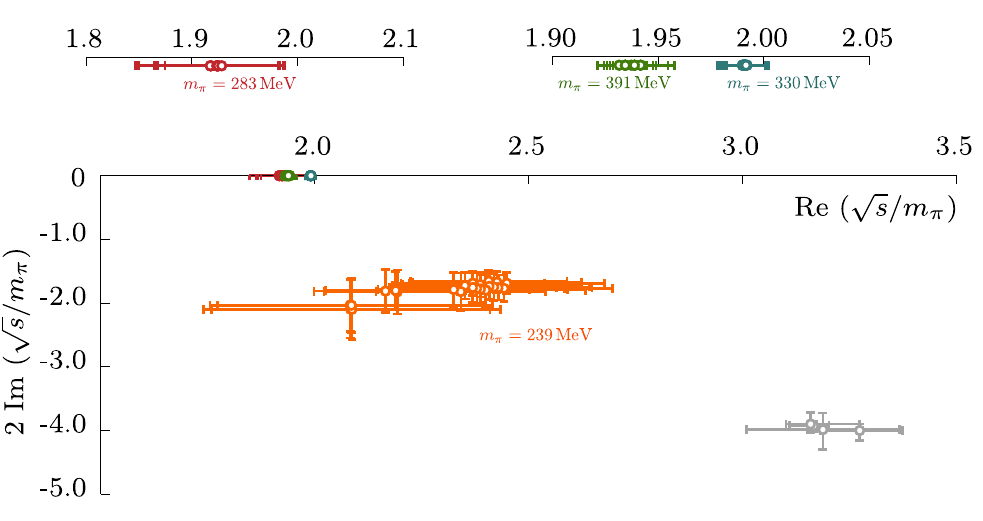}
\caption{
$\sigma$ pole evolution with changing light-quark mass corresponding to $m_\pi\sim 391$ MeV (green, bound-state, non-dispersive, taken from Ref.~\cite{Briceno:2016mjc}), 330 MeV (blue, bound-state, non-dispersive, taken from Ref.~\cite{Rodas:2023gma}), 283 MeV (red, virtual bound-state, dispersive), 239 MeV (orange, resonance, dispersive). Gray points correspond to dispersive extractions from experimental data given in Refs.~\cite{Caprini:2005zr,Garcia-Martin:2011nna,Moussallam:2011zg}. }\label{fig:evolution}
\end{figure*}

In the $m_\pi \sim 283$ MeV case (Fig.~\ref{fig:sigma_poles856}), the $S0$ lattice amplitudes indicated that the $\sigma$ could be either a virtual bound-state\footnote{a pole on the real energy axis below threshold on the unphysical Riemann sheet.} \emph{or} a subthreshold resonance, depending upon parameterization choice, with a spread in pole positions. Those dispersed amplitudes that meet the metric cuts show a reduced scatter in pole location. As described in Ref.~\cite{Rodas:2023gma}, when the complex conjugate pole pair of a subthreshold resonance meet on the real axis below threshold the pole location's dependence on the amplitude parameters develops an infinite slope. Near this point, the slopes are large, causing small uncertainties on the parameters to become large uncertainties on the pole location. We do not plot the dispersive pole locations for these noisy results but merely comment that they are compatible with the plotted virtual bound-state cases. In the current analysis, a definitive statement about whether the state is a virtual bound-state, or a subthreshold resonance at this pion mass cannot be made.
The $\pi\pi$ couplings of these plotted poles are also presented in the figure, where we see that the dispersive results have larger statistical uncertainties but reduced systematic spread with respect to the conventional analyses presented in Refs.~\cite{Briceno:2016mjc,Rodas:2023gma}. 

 In our $m_\pi \sim 283$ MeV analysis, the selected dispersed central values find no real Adler zero for $S0$. In those cases, a very noisy third pole, companion to the pair of $\sigma$ poles, appears close to the left-hand cut in the unphysical Riemann sheet. 

In this work, we have considered two pion masses for which the $\sigma$ appears as a pole in the unphysical Riemann sheet, while at higher pion masses lattice calculations indicate that the $\sigma$ is a bound-state pole on the physical Riemann sheet. 
In these latter cases, a dispersive analysis is typically not required to determine accurately the $\sigma$ pole location, as it is tightly constrained by finite-volume energy levels lying close to the pole. Nevertheless, it is possible to construct applicable dispersion relations by explicitly including the $\sigma$ in $T^0(s,t,u)$ as ``fixed'' poles in $s$ and $u$. The $s$--channel pole remains a pole when the amplitude is projected into the $s$--channel $S$--wave, while the $u$--channel pole generates a cut that is present in all partial-waves. 
Recently,  Ref.~\cite{Cao:2023ntr} applied dispersion relations in an approach different to that explored in this paper, in a case where a bound-state $\sigma$ is present using lattice data at $m_\pi\sim 391$ MeV previously published by {\it hadspec}, finding a $\sigma$ pole compatible with the undispersed analysis in Ref.~\cite{Briceno:2017qmb}.

\vspace{3mm}

Finally, taking the now robust $\sigma$ pole results from dispersive analysis at $m_\pi \sim 239,\, 283$ MeV and supplementing them with two heavier quark masses where dispersive analysis is not required, we show in \cref{fig:evolution} the evolution with changing quark mass of the $\sigma$ pole. In distinction to the narrow $P$-wave $\rho$ resonance, which exhibits a simple quark-mass dependence corresponding to an approximately constant coupling to $\pi\pi$~\cite{Rodas:2023gma,Hanhart:2008mx,Hanhart:2014ssa}, the $S$-wave $\sigma$ undergoes a rapid transition from bound to resonant state, appearing to pass through a virtual bound-state stage in a narrow region of pion mass.

\section{Summary} 
\label{sec:summary}

We have presented a dispersive approach to analyze elastic hadron-hadron scattering information provided by lattice QCD, applied here to the case of $\pi\pi$ scattering at two pion masses. We have observed that the sensitivity of $S$--wave scattering lengths and broad resonance pole locations to choice of parameterization form can be largely eliminated.  The method also yields reliably the scattering amplitudes below threshold, and in this region we have pinned down the location of amplitude zeroes (Adler zeroes) when they appear. This indicates that reliable results can be extracted from lattice QCD calculations without the need to \emph{enforce} amplitude features, such as Adler zeroes, not directly motivated by the lattice data.
The now well-constrained amplitudes obtained in this study can be used in future lattice QCD calculations in which the $\pi\pi$ $S$-wave scattering system is coupled to external currents~\cite{Lellouch:2000pv,Briceno:2014uqa,Briceno:2015tza,Briceno:2022omu}, where well-defined \emph{form-factors} of resonances can be extracted from pole residues, providing structural information that will aid in the determination of the compositeness of the $\sigma$ as the light quark mass is varied.

\acknowledgments
We thank our colleagues within the Hadron Spectrum Collaboration.
   
	AR, JJD, and RGE acknowledge support from the U.S. Department of Energy contract DE-AC05-06OR23177, under which Jefferson Science Associates, LLC, manages and operates Jefferson Lab. 
	AR, JJD also acknowledge support from the U.S. Department of Energy award contract DE-SC0018416. This work was done as part of the ExoHad Topical Collaboration.

The software codes
{\tt Chroma}~\cite{Edwards:2004sx}, {\tt QUDA}~\cite{Clark:2009wm,Babich:2010mu}, {\tt QUDA-MG}~\cite{Clark:SC2016}, {\tt QPhiX}~\cite{ISC13Phi}, and {\tt QOPQDP}~\cite{Osborn:2010mb,Babich:2010qb}, and {\tt Redstar}~\cite{Chen:2023zyy} were used. 
The authors acknowledge support from the U.S. Department of Energy, Office of Science, Office of Advanced Scientific Computing Research and Office of Nuclear Physics, Scientific Discovery through Advanced Computing (SciDAC) program.
Also acknowledged is support from the U.S. Department of Energy Exascale Computing Project.
The contractions were performed on clusters at Jefferson Lab under the USQCD Initiative and the LQCD ARRA project. This research was supported in part under an ASCR Leadership Computing Challenge (ALCC) award, and used resources of the Oak Ridge Leadership Computing Facility at the Oak Ridge National Laboratory, which is supported by the Office of Science of the U.S. Department of Energy under Contract No. DE-AC05-00OR22725.
This research is also part of the Blue Waters sustained-petascale computing project, which is supported by the National Science Foundation (awards OCI-0725070 and ACI-1238993) and the state of Illinois. Blue Waters is a joint effort of the University of Illinois at Urbana-Champaign and its National Center for Supercomputing Applications. This research used resources of the National Energy Research Scientific Computing Center (NERSC), a DOE Office of Science User Facility supported by the Office of Science of the U.S. Department of Energy under Contract No. DE-AC02-05CH11231.
The authors acknowledge the Texas Advanced Computing Center (TACC) at The University of Texas at Austin for providing HPC resources. Gauge configurations were generated using resources awarded from the U.S. Department of Energy INCITE program at Oak Ridge National Lab, and also resources awarded at NERSC.

\appendix

\section{Dispersive inputs}
\label{app:dr_input}

As described in the text, when implementing~\cref{eq:DR} the input $\mathrm{Im}\, t^{I'}_{\ell'}(s')$ is needed over the entire energy region from threshold to infinity. In practice, we split this integral into two pieces, below $s_h$ where lattice-obtained partial-wave amplitudes are used, and above $s_h$, where Regge-like parameterizations are used. Due to the polynomial suppression of the integral kernels at high-energies, the sensitivity of dispersed amplitudes at low energies to the details of the amplitudes above $s_h$ is weak.

\subsection{Lattice obtained partial-wave input}
\label{app:pws}

The partial-wave lattice inputs are described by using parameterizations presented in Refs.~\cite{Briceno:2016mjc,Rodas:2023gma}, parameterizations fitted to the results presented in Ref.~\cite{Wilson:2015dqa}, or parameterizations fitted to the results presented in~\cref{app:860I2}. For a detailed summary of the functional forms used, see~\cref{app:func_forms}. They are constrained as follows:

$\mathbf{S2, D2}$: these amplitudes, whose spectra are compatible with there being no inelasticity, are constrained by fitting energy levels up to around $a_t E_\mathsf{cm} = 0.22$ for both $m_\pi \sim 239, \,283$ MeV masses. They describe weak repulsive non-resonant scattering.

 $\mathbf{P1}$: this amplitude is constrained by fitting energy levels up to around $a_t E_\mathsf{cm} = 0.185(0.19)$ for $m_\pi \sim 239 (283)$ MeV, very slightly above the $K\bar{K}$ threshold. As presented in Ref.~\cite{Wilson:2015dqa}, there is negligible inelasticity up to $a_t E_\mathsf{cm} = 0.22$, and as such, we use an extrapolation of the determined elastic amplitude up to $s_h$. This partial wave is dominated by the $\rho$ resonance in the elastic region, which can be fitted well by functional forms with few free parameters, such as a Breit-Wigner form. These amplitudes have a smooth featureless extrapolation up to $s_h$. The fact that the phase-shift is very close to $180^\circ$ in the inelastic region means the imaginary part that enters the dispersion relations is very close to zero.

 $\mathbf{S0}$: this amplitude is constrained by fitting energy levels up to around $a_t E_\mathsf{cm} = 0.1585(0.16)$ for $m_\pi \sim 239 (283)$ MeV, slightly below the $K\bar{K}$ threshold. The amplitudes feature slow energy dependence in the elastic region. The presence of a scalar resonance analogous to the $f_0(980)$ can cause the $K\bar{K}$ amplitude to turn on rapidly, and we have not yet attempted coupled-channel descriptions in this region, where $4\pi$ channels may also be relevant. In practice, we use extrapolations of the elastic amplitude up to $s_h$, and for different parameterizations, these extrapolations can differ significantly. However, as observed in Figures~\ref{fig:roy_select860}, \ref{fig:roy_select856}, the main variation in the dispersion relation metrics comes from the behavior of the amplitudes at threshold (characterized by the scattering length), so it appears that there is little sensitivity to this extrapolation region, suppressed as it is by the rapidly decreasing dispersion relation kernels.

\begin{figure*}[htb]
\includegraphics[width=1.0\textwidth]{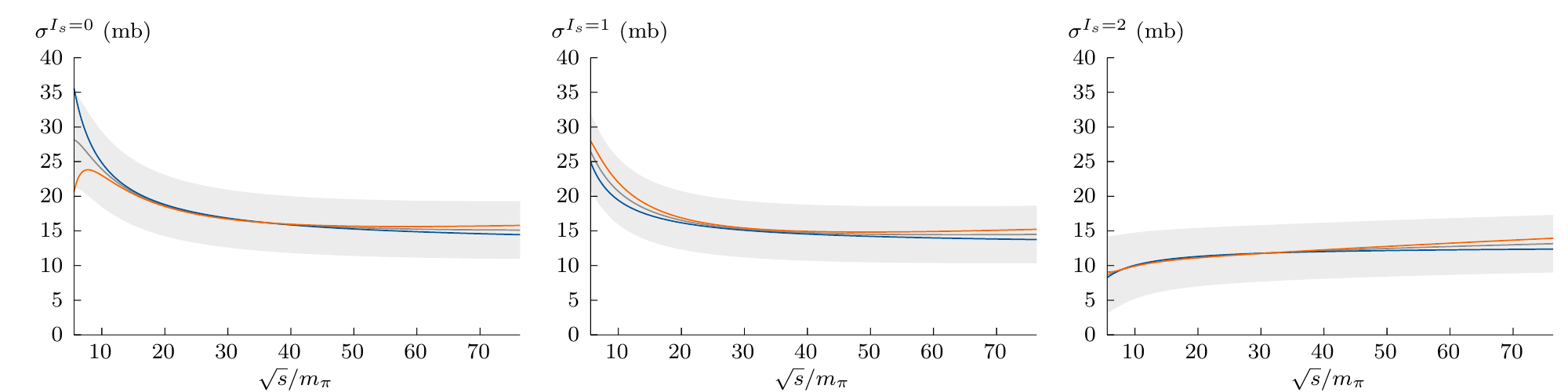}
\caption{ Regge contributions to $\sigma^{I_s=0}(s)$ (left), $\sigma^{I_s=1}(s)$ (center) and $\sigma^{I_s=2}(s)$ (right) for $m_\pi \sim 239$ MeV. Parameterizations taken from Ref.~\cite{Garcia-Martin:2011iqs} (blue) and Ref.~\cite{Caprini:2011ky} (orange), scaled to be applicable to the current pion mass. In our dispersive analysis, we use the average of the two (gray) assigning a conservative 30\% fractional uncertainty to each $I_t$ amplitude.
}\label{fig:ReggeAmps}
\end{figure*}

\begin{figure*}[htb]
\includegraphics[width=1.0\textwidth]{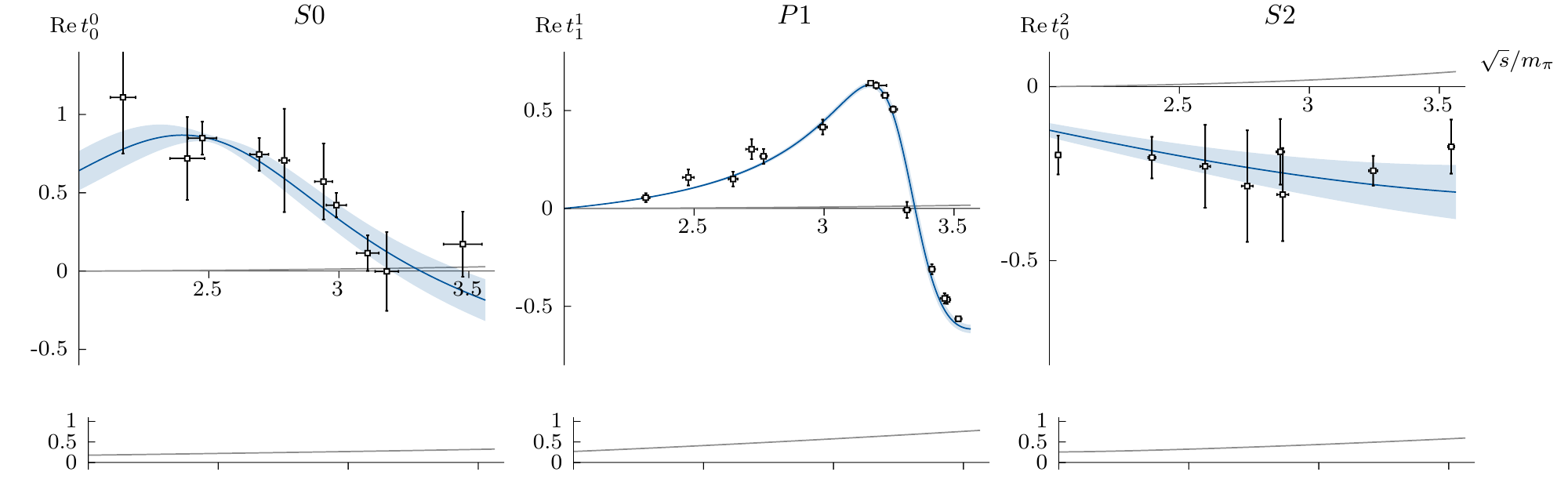}
\caption{ 
{ Top row: Blue curves show the real part of the $S0$, $P1$, $S2$ partial-wave amplitudes for $m_\pi \sim 239$ MeV (amplitude combination indicated by the black star in \cref{fig:metric_cuts}). The gray curves lying close to zero in each case show the contribution to the amplitude, $\tilde{f}^I_\ell(s) \big|_\mathrm{Regge}$, coming from the high-energy amplitude parameterization (as defined in \cref{fRegge}).  \\
Bottom row: Ratio, in absolute value, of $\tilde{f}^I_\ell(s) \big|_\mathrm{Regge}$ and the denominator in the parenthesis of~\cref{eq:dsq}, indicating the high-energy contribution in units of the relative uncertainty at each energy.  
} 
}\label{fig:ReggeContrib}
\end{figure*}

\subsection{High-energy parameterization}
\label{app:regge}

Generally, it is observed that scattering amplitudes at high energy cease to show features of individual resonances, becoming smooth functions of energy. In the limit $s \gg t$, the \emph{complex angular momentum}, or ``Regge'' approach proves to be an efficient method to parameterize scattering amplitudes in terms of a modest number of parameters describing $t$--channel Regge trajectories and their residues~\cite{Veneziano:1968yb,Lovelace:1968kjy,Shapiro:1969km}. This approach has been used extensively to describe experimental scattering data, with good descriptions of $\pi\pi$ scattering at high energy being obtained by inclusion of Pomeron, $\rho$ and $f_2$ trajectories~\cite{Garcia-Martin:2011iqs,Caprini:2011ky,Pelaez:2003ky}. 

For the current analysis at unphysical values of the light quark mass, we lack any high-energy scattering data with which to constrain a Regge parameterization, but we can use the relationship between Regge trajectories and resonance states exchangeable in the $t$--channel to infer the required quark-mass scaling. In particular, we will use an average of the parameterizations given in Refs.~\cite{Garcia-Martin:2011iqs,Caprini:2011ky} (specifically we use the ``CFD" results from Ref.~\cite{Garcia-Martin:2011iqs}), adapted to our pion masses in a simplistic approach.  These formulas provide us with a description for the three different isospins in the $t$-channel exchanges $I_t=0,\,1,\,2$. 

The Pomeron trajectory is typically associated with gluonic exchanges, and as such we will assume that the parameters obtained from fits to physical data can be used without adjustment for the changed light quark mass. We will also assume that the $I_t=2$ Regge trajectory, which plays a very minor role, does not depend on the quark mass. On the other hand, the $\rho$ trajectory includes the $\rho$ resonance, and the residue of this trajectory in $\pi \pi$ scattering can be related to the decay width for $\rho \to \pi \pi$, which does change with varying light quark mass. We keep the Regge trajectory $\alpha_\rho(s)$ fixed at the physical value, and only adjust the residue for the changed quark mass. Equation (E.5) from Ref.~\cite{Ananthanarayan:2000ht},
\begin{equation}
\Gamma_\rho=\frac{\lambda}{96 \pi M_\rho^2}\left(M_\rho^2-4 M_\pi^2\right)^{\frac{3}{2}},
\end{equation}
indicates how to scale the residue (which is proportional to $\lambda$) given the $M_\rho, \Gamma_\rho$ computed in lattice QCD at an unphysical light quark mass. For $m_\pi\sim 239$ MeV we take $M_\rho=793.7$ MeV and $\Gamma_\rho=91.2$ MeV, while for $m_\pi\sim 283$ MeV we use $M_\rho=798.6$ MeV and $\Gamma_\rho=60.5$ MeV. The residue, $\beta$, present for $\rho$ exchange, given in Refs.~\cite{Garcia-Martin:2011iqs,Caprini:2011ky}, is scaled by the ratio of the lattice $\lambda$ to the physical value, which we take as $\lambda_\mathrm{phys}=72$. The $\rho$--computed ratio is also used (as a crude first approximation) to scale the $f_2$ residue.

The Regge implementations of Refs.~\cite{Garcia-Martin:2011iqs,Caprini:2011ky}, with the quark-mass scaled residues, are used to calculate the $\pi \pi$ cross-sections in the $s$-channel, according to
\begin{equation}
\sigma^{I_s}(s)=\frac{\mathrm{Im}\,  T^{I_s}(s,0)}{\sqrt{s(s-4m_\pi^2)}}.
\end{equation}
Figure~\ref{fig:ReggeAmps} shows these for the $m_\pi \sim 239$ MeV case, along with the average of the two parameterizations, with a conservatively assigned, uncorrelated $30\%$ error for each $I_t$ total Regge amplitude contribution, which we propagate through our dispersive analysis.

These amplitudes are used in the dispersion relations from $s_h$ up to infinity, but note that their contributions are heavily suppressed by the kernels, which as shown in Figure~\ref{fig:RoyKernels}, fall rapidly with increasing energy. In \cref{fig:ReggeContrib}, we present the contribution of the high-energy parameterized amplitude to the dispersed amplitudes in the elastic scattering region. We define
\begin{equation}
\tilde{f}^I_\ell(s) \big|_\mathrm{Regge} \equiv \sum_{I', \ell'} \int_{s_h}^\infty \!\!ds' \; \mathrm{Re}\, K^{II'}_{\ell\ell'}(s',s) \; \mathrm{Im} \, t^{I'}_{\ell'}(s') \big|_\mathrm{Regge} \, ,
\label{fRegge}
\end{equation}
and compare this to the total dispersed amplitude, which is dominated by the amplitudes constrained by lattice QCD data for $s < s_h$. The plotted ratio between the high-energy contribution and the relative uncertainty, $\Delta \big[ \tilde{f}^I_\ell(s) - f^I_\ell(s)\big]$, indicates that the results of the analysis presented above are not sensitive to the detailed modeling of the Regge amplitudes.

\section{Dispersed amplitude metrics}
\label{app:extended_values}

\begin{figure}[!b]
  \includegraphics[width=0.5\textwidth]{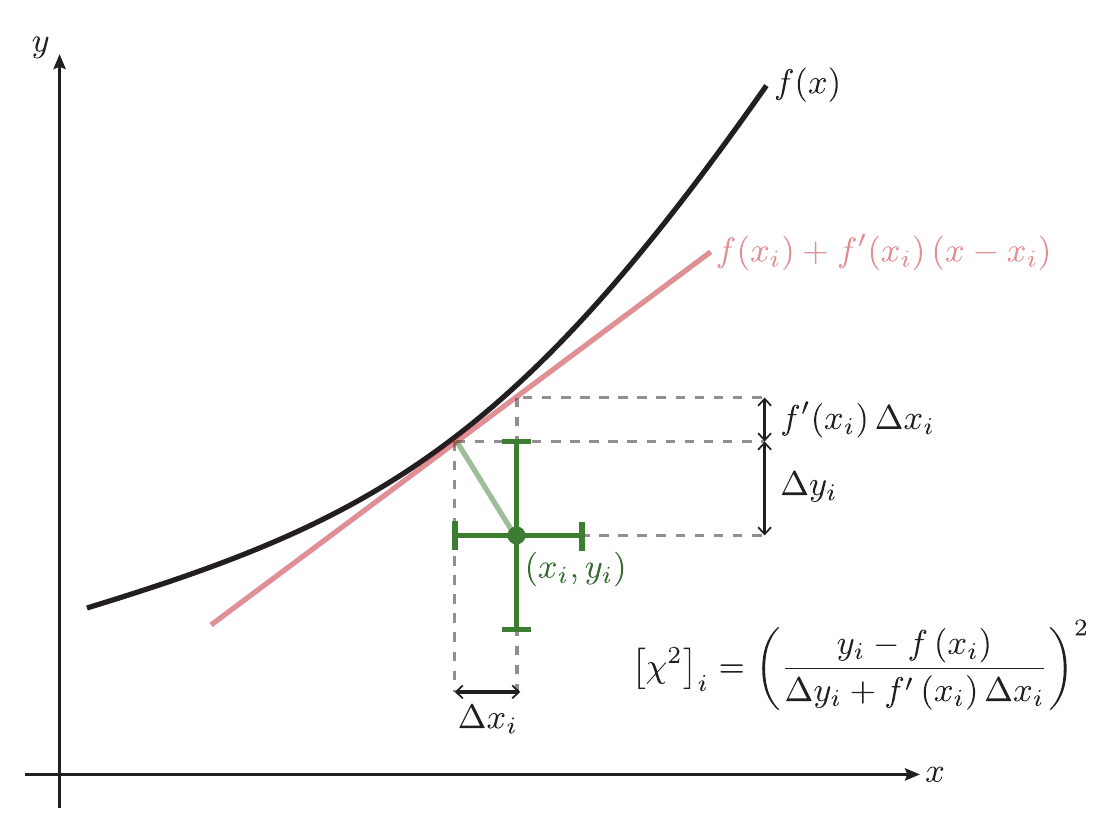}
  \caption{Geometric illustration of the construction of a linearized $\chi^2$ contribution for a data-point having $x,y$ correlation of $-1$. The data-point shown will contribute $1$ to the $\chi^2$ in agreement with the apparent $1\sigma$ deviation from the curve along the diagonal direction. }
  \label{fig:chisq_geom}
\end{figure}

As discussed in the manuscript, we retain only those dispersed amplitudes which are both compatible with unitarity, determined by having real part in agreement with the input amplitude (which itself was exactly unitarity preserving), and which provide a reasonable description of the original lattice data. These criteria are assessed using two numerical metrics.

A metric which compares the real part of the dispersed amplitude, $\tilde{f}^I_\ell(s) \equiv \mathrm{Re}\, \tilde{t}^I_\ell(s)$, to the real part of the input amplitude, $f^I_\ell(s) \equiv \mathrm{Re}\, t^I_\ell(s)$, in order to enforce unitarity, is
\begin{equation}
\big[ d^2 \big]^I_\ell \equiv \sum_{i=1}^{N_\mathrm{smpl}} 
\left( 
   \frac{\tilde{f}^I_\ell(s_i) - f^I_\ell(s_i) }
   {\Delta \big[ \tilde{f}^I_\ell(s_i) - f^I_\ell(s_i)  \big]} 
\right)^2 \, . 
\end{equation}

The sampling of points in the sum runs from just above the $\pi\pi$ threshold\footnote{In practice, this difference is calculated at 100 evenly spaced energy values in the range between the two-pion threshold and $a_t \sqrt{s} =0.14$, of which points above $2a_t m_\pi+0.005$ are used to obtain the $d^2$ value. The energy of the first point used is around $a_t \sqrt{s} = 0.0841(0.0995)$ for $m_\pi\sim239(283)$ MeV.}. Excluding the lowest 30 MeV above threshold reduces the sensitivity to high-order derivatives of the scattering amplitude at threshold, a sensitivity that is peculiar to this particular difference. In total, we use 91 equally spaced points for \mbox{$m_\pi \sim 239$ MeV} and 89 points for $m_\pi \sim 283$ MeV in the evaluation of this metric. The uncertainty on the difference at each energy sample (appearing in the denominator) is calculated by linear propagation of the (correlated) amplitude parameter uncertainties and the (uncorrelated) high-energy Regge model uncertainty.

\begin{SCfigure*}[][!ht]
\includegraphics[width=0.7\textwidth]{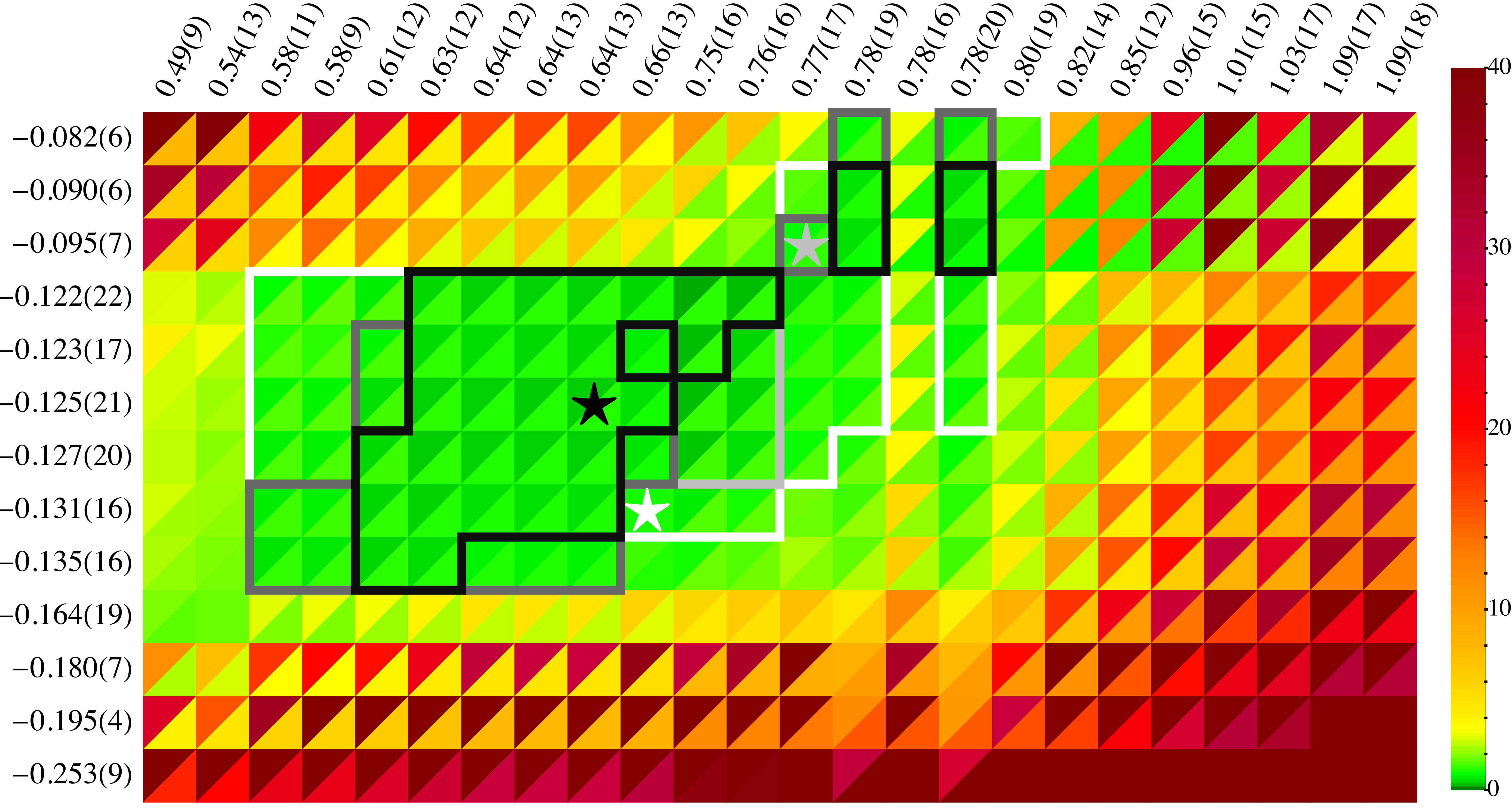}
\caption{For $m_\pi \sim 239$ MeV, for each combination of parameterizations of $S0$ (columns) and $S2$ (rows), each ordered by the magnitude of their scattering length in $m^{-1}_\pi$ units, the box is colored according to the average value of $d^2/N_\mathrm{smpl}$ (upper triangle) and $\tilde{\chi}^2/N_\mathrm{lat}$ (lower triangle) over $S0$, $P1$, $S2$. The different regions indicated by the color contours show the parameterization combinations respecting the cuts described in the text. Three example parameterization combinations are indicated by black, gray, and white stars, which are used for illustration in other figures.
}\label{fig:metric_cuts}
\end{SCfigure*}

\begin{figure*}[!hbt]
\includegraphics[width=0.9\textwidth]{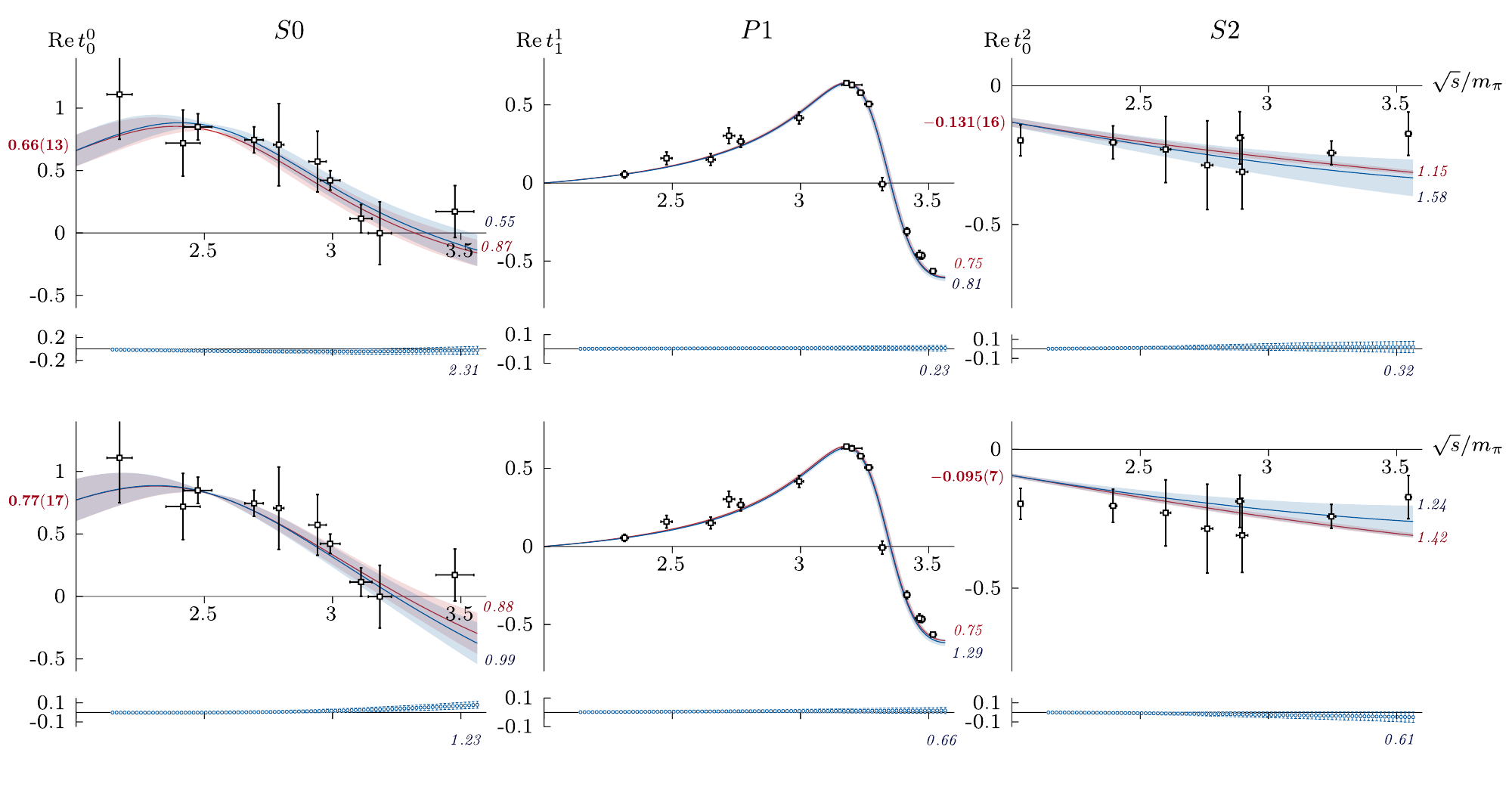}
\caption{For $m_\pi \sim 239$ MeV, real parts of $S0$, $P1$, $S2$ partial-wave amplitudes for input (red) and dispersed output (blue). Amplitude combinations correspond to the stars in \cref{fig:metric_cuts}: white (upper panel), gray (lower panel). Labels as described in the caption of Figure~\ref{fig:amp860}.
}\label{fig:metric_cuts_alt}
\end{figure*}

\vspace{3mm}
In order to compare the dispersed amplitudes directly to the `lattice amplitude data', we use a $\chi^2$-like construction which accounts for both the correlation between different lattice data points and the correlation between the amplitude uncertainty and the energy uncertainty of each point,
\begin{align}
\big[ \tilde{\chi}^2 \big]^I_\ell &\equiv \sum_{i,j=1}^{N_\mathrm{lat}} 
\left( \tfrac{ \mathfrak{f}_i - \tilde{f}^I_\ell(s_i)}{\Delta_i} \right)
\mathrm{corr}(\mathfrak{f}_i, \mathfrak{f}_j)^{-1}
\left( \tfrac{ \mathfrak{f}_j - \tilde{f}^I_\ell(s_j)}{\Delta_j} \right) \, , \nonumber\\
\Delta_i^2 &= \big( \Delta \mathfrak{f}_i \,\,\, \frac{d \tilde{f}^{I}_\ell(s_i)}{d E_i} \Delta E_i \big) \begin{pmatrix}1 & -c_i\\ -c_i & 1 \end{pmatrix} \begin{pmatrix} \Delta \mathfrak{f}_i \\ \frac{d \tilde{f}^{I}_\ell(s_i)}{d E_i}\Delta E_i  \end{pmatrix},
\end{align}
In the second expression, the ``$x,y$'' correlation, $c_i$, for points obtained from a linearized, L\"uscher finite-volume analysis is $\pm 1$, and in \cref{fig:chisq_geom} we present a geometric illustration of the construction of the effective uncertainty, compatible with the equation above, for the case $c_i = -1$. $\Delta^2_i$ is an ``effective variance weight"; explicit examples with and without correlations between the $x$ and $y$ values, and a justification of the formula above are given in Refs.~\cite{Jefferys:1980AJ,Lybanon:1982}\footnote{The methods described in Refs.~\cite{Jefferys:1980AJ,Jefferys:1981AJ,Lybanon:1982} present a solution to fitting data with errors in both $x,\,y$ variables, for which the $x_i$ values are iterated, with the suggested starting point $x_i=E_i$ for our case.}. The values of $c_i$ and $\mathrm{corr}(\mathfrak{f}_i, \mathfrak{f}_j)$ are obtained from the original lattice ensemble distributions\footnote{A few data points appear very close to local maxima or minima of $\tilde{f}^I_\ell(s_i)$, for which the linearized statistical uncertainty on $\mathfrak{f}_i$ and/or the derivative $\frac{d \tilde{f}^{I}_\ell(s_i)}{d E_i}$ are almost zero. In these cases, the error is underestimated. For these problematic $\mathfrak{f}_i$ points, we took the maximum statistical or systematic error, when varying the energy level $E_i$ within uncertainties. For $\tilde{f}^{I}_\ell(s_i)$, the second order correction $\frac{d^2 \tilde{f}^{I}_\ell(s_i)}{d^2 E_i} (\Delta E_i)^2/2$ is sizeable for some points (over $30\%$ of $|\frac{d \tilde{f}^{I}_\ell(s_i)}{d E_i} \Delta E_i|$), and in those cases we substituted $\frac{d \tilde{f}^{I}_\ell(s_i)}{d E_i} \Delta E_i$ by either $\frac{d \tilde{f}^{I}_\ell(s_i)}{d E_i} \Delta E_i+\frac{d^2 \tilde{f}^{I}_\ell(s_i)}{d^2 E_i} (\Delta E_i)^2/2$ or $\frac{d \tilde{f}^{I}_\ell(s_i)}{d E_i} \Delta E_i-\frac{d^2 \tilde{f}^{I}_\ell(s_i)}{d^2 E_i} (\Delta E_i)^2/2$, selecting the term that produces the larger value for $\Delta^2_i$.}.

Applying these metrics to the amplitudes obtained from the dispersion relations generates Figures~\ref{fig:roy_select860}, \ref{fig:roy_select856}. In the manuscript we apply cuts of $d^2/N_\mathrm{smpl} < 1$ and ${\tilde{\chi}^2/N_\mathrm{lat} < 2}$, retaining only those amplitudes which satisfy these for all partial-waves, $S0,P1,S2$. In \cref{fig:metric_cuts}, we show the effect of relaxing these cuts somewhat. The four superimposed boundaries correspond to:
\begin{itemize}
\item[]{\bf white}:  amplitude combinations with all ${d^2/N_\mathrm{smpl} < 3}$ and $\tilde{\chi}^2/N_\mathrm{lat} < 3$ 
 \item[]{\bf gray}:  amplitude combinations with all ${d^2/N_\mathrm{smpl} < 1}$ and $\tilde{\chi}^2/N_\mathrm{lat} < 3$
 \item[]{\bf dark gray}: amplitude combinations with all ${d^2/N_\mathrm{smpl} < 2}$ and $\tilde{\chi}^2/N_\mathrm{lat} < 2$ 
  \item[]{\bf black}: amplitude combinations with all ${d^2/N_\mathrm{smpl} < 1}$ and $\tilde{\chi}^2/N_\mathrm{lat} < 2$  \, ,

\end{itemize}
and the figure indicates that somewhat looser cuts can \emph{slightly} increase the range of acceptable scattering length values, but as shown in \cref{fig:metric_cuts_alt} (for the sample points indicated by the white and gray stars in \cref{fig:metric_cuts}), they do so by allowing increasing departures from unitarity, or from an acceptable description of the lattice data.

\begin{SCfigure*}
\includegraphics[width=.6\textwidth]{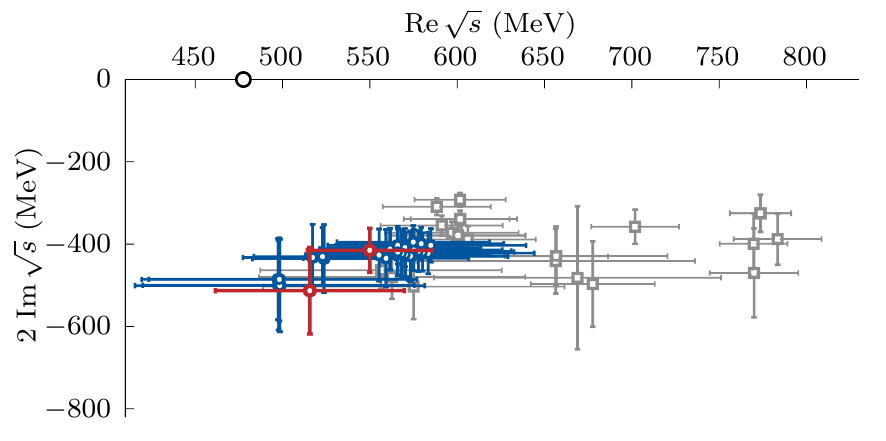}
\caption{For $m_\pi \sim 239$ MeV, $S0$ pole locations for lattice-fit amplitudes (gray), dispersed amplitudes satisfying the tighter metric cuts described in the text (blue), and for the two sample points (white and gray stars in \cref{fig:metric_cuts}) satisfying looser cuts (red).
}\label{fig:POLES_MULTI}
\end{SCfigure*}

Nevertheless, if one allows these looser cuts, we note that the $\sigma$ pole location remains within the region established by the tighter cuts, as shown in \cref{fig:POLES_MULTI}, indicating that our particular choice of metric cuts is not introducing a significant systematic error.

\section{Minimally subtracted dispersion relations (``GKPY'')}
\label{app:GKPY}

\begin{figure*}[!tbh]
\includegraphics[width=0.48\textwidth]{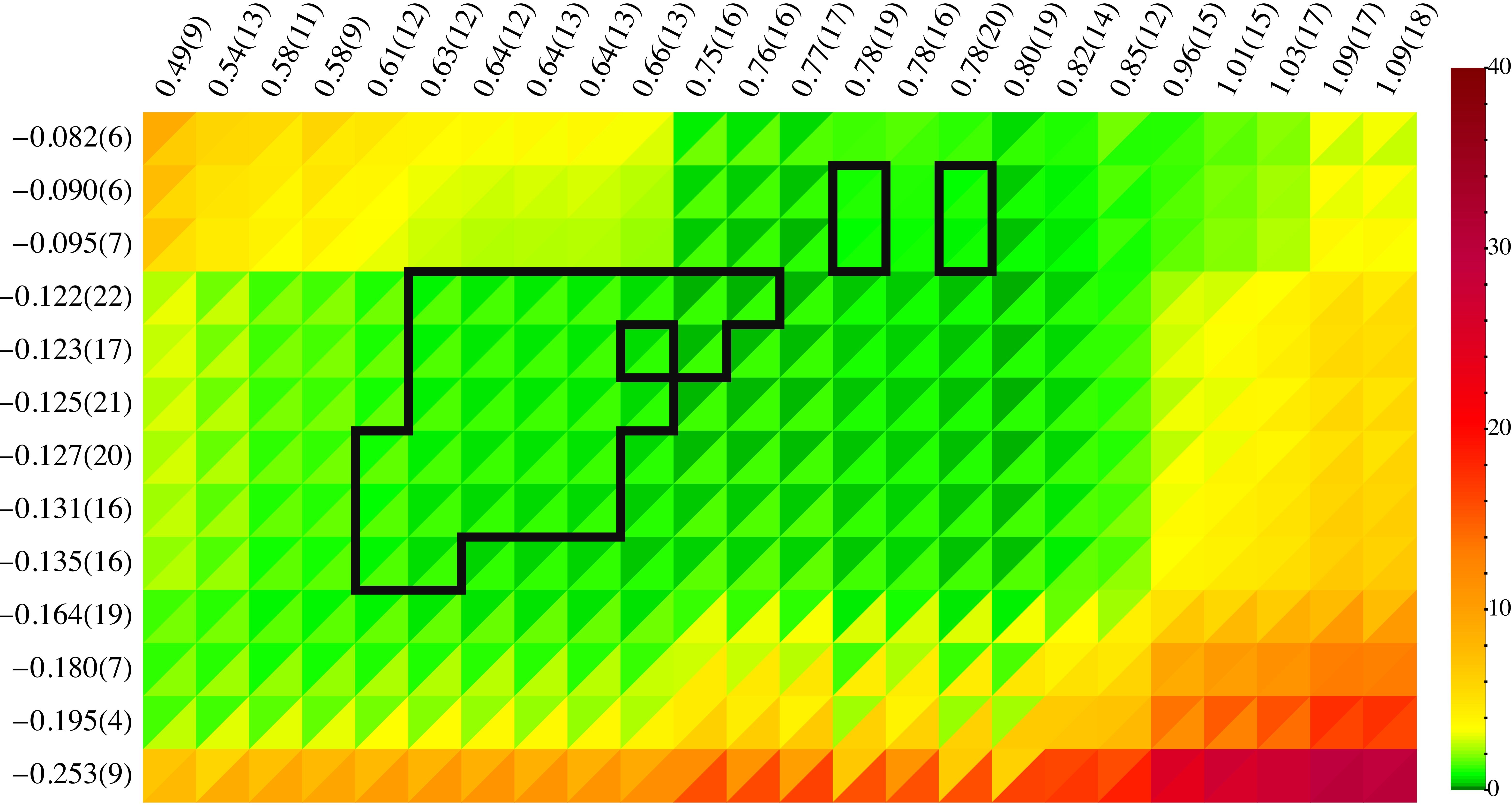} 
\hspace{2mm}
\includegraphics[width=0.48\textwidth]{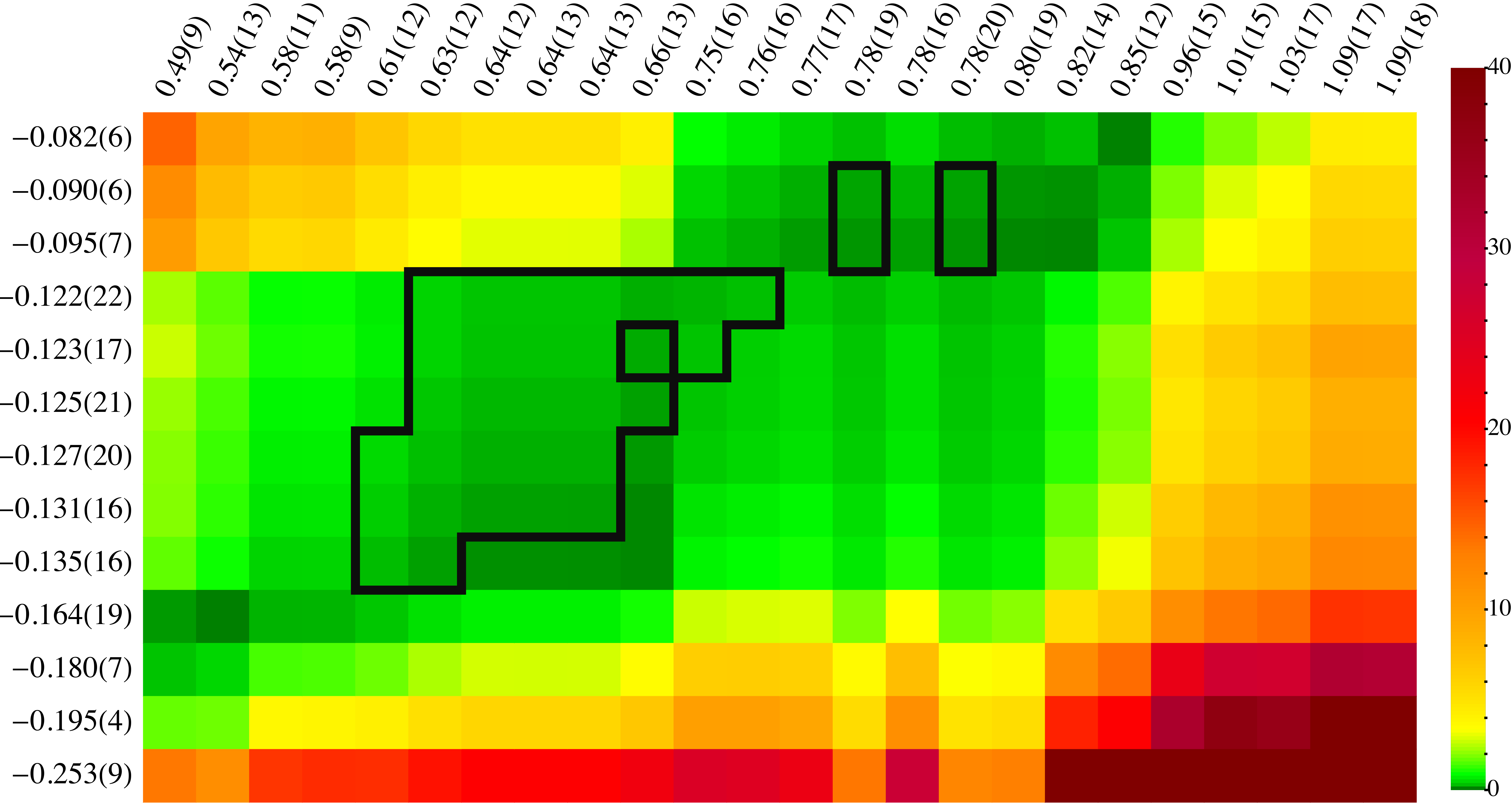}
\caption{ For $m_\pi \sim 239$ MeV, left panel shows average values of $d^2/N_\mathrm{smpl}$ (upper triangle) and $\tilde{\chi}^2/N_\mathrm{lat}$ (lower triangle) for each pair of $S0$ (columns) and $S2$ (rows) parameterizations, computed using \emph{minimally-subtracted} dispersion relations. The black outline indicates those amplitudes that passed the metric cuts in the \emph{twice-subtracted} case. Right panel shows values of $\left(\frac{\psi}{\Delta \psi}\right)^2$ as defined in the text, testing the degree to which the Olsson sum rule is satisfied.
}\label{fig:GKPY860}
\end{figure*}

\begin{figure*}[!tbh]
\includegraphics[width=0.48\textwidth]{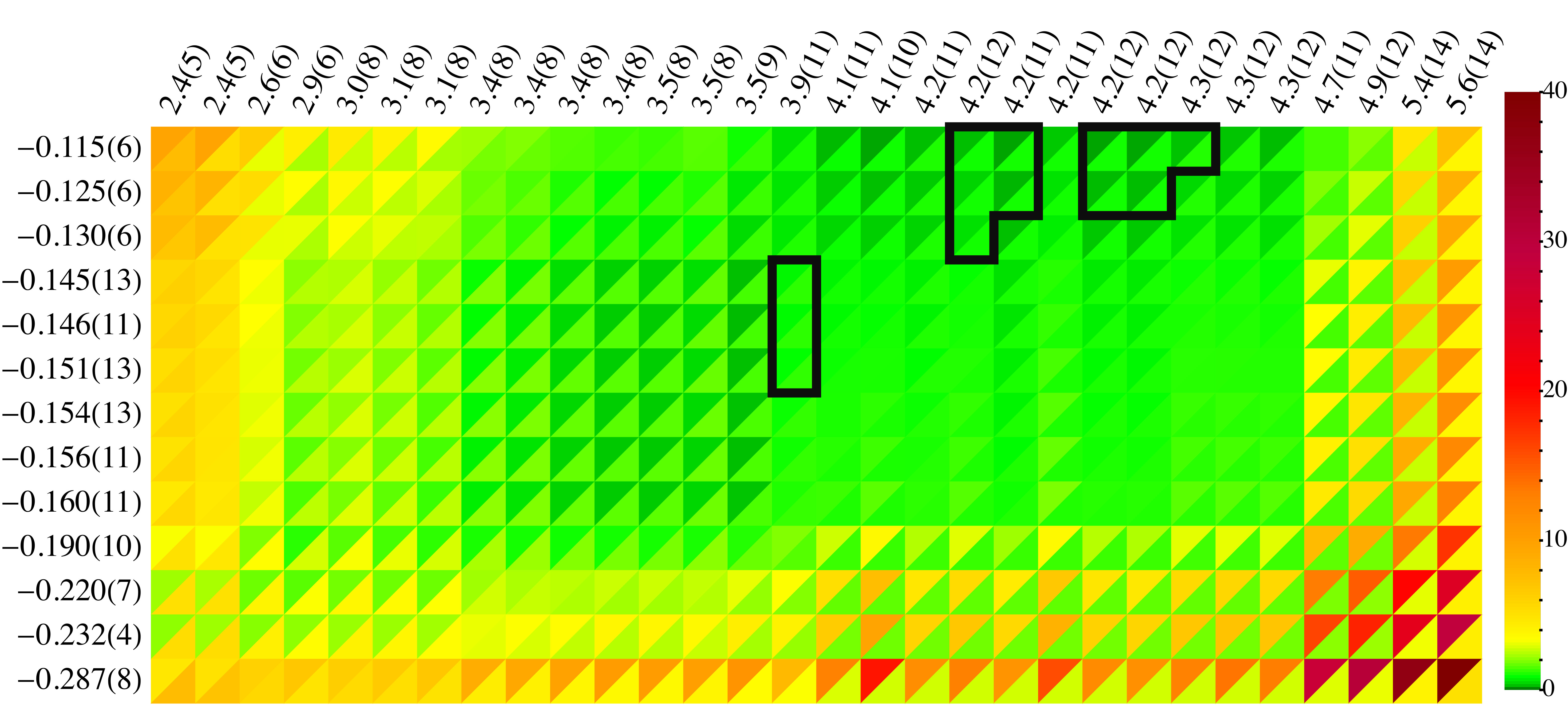}
\hspace{2mm}
\includegraphics[width=0.48\textwidth]{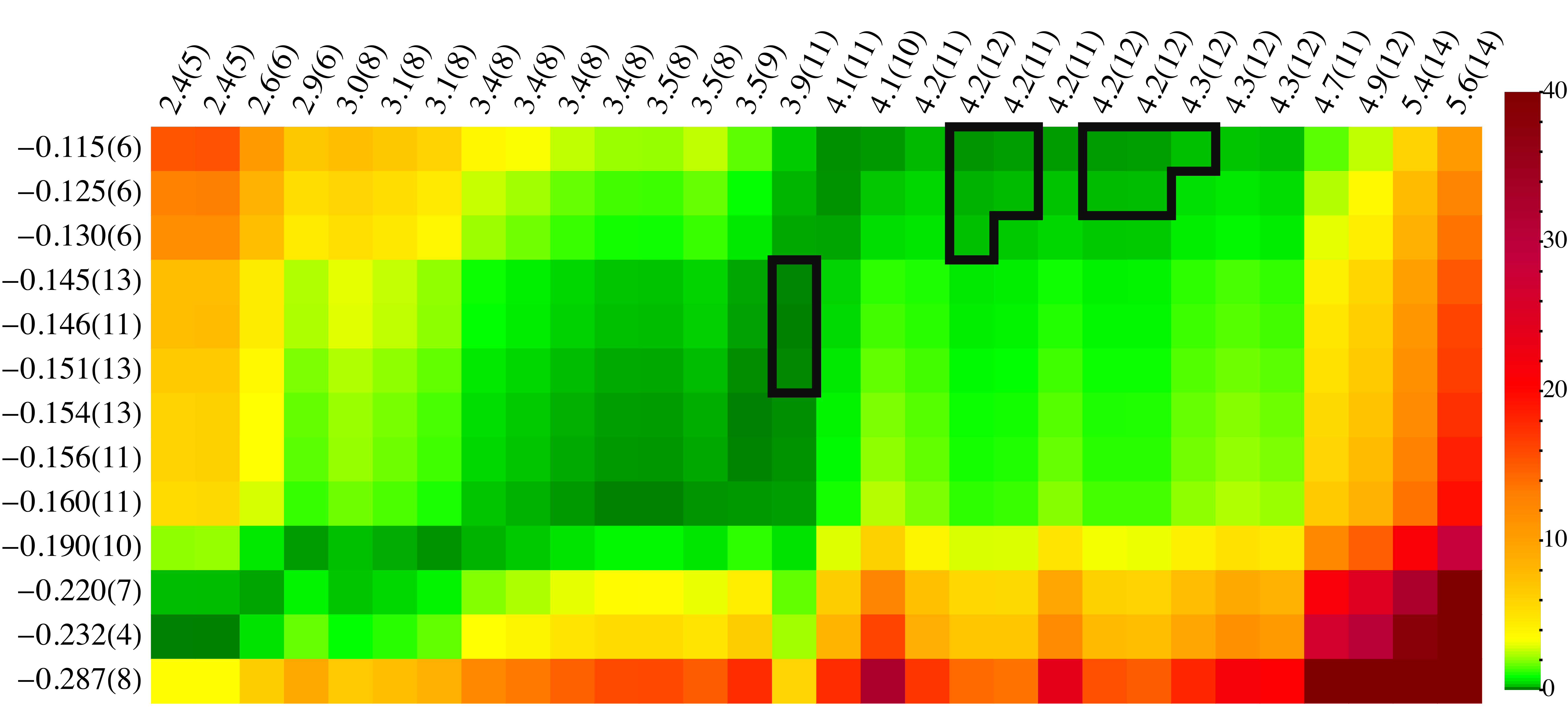}   
\caption{As Figure~\ref{fig:GKPY860}, but for $m_\pi \sim 283$ MeV.
}\label{fig:GKPY856}
\end{figure*}

\emph{Minimally-subtracted} dispersion relations, often referred to as GKPY equations, are constructed making only those subtractions \emph{required} to get convergence, namely one subtraction~\cite{Garcia-Martin:2011iqs}. The smaller number of subtractions means the high-energy contributions to the dispersion relations are not as strongly suppressed, with the kernel functions, in this case, falling off as $s^{-2}$ rather than $s^{-3}$ as was the case for \emph{twice-subtracted} (``Roy'') dispersion relations. The subtraction functions in this case are just constants which, as before, depend only on the $S$-wave scattering lengths,
\begin{align}
 \tau^0_0(s) / m_\pi &= \tfrac{1}{3} \big( a^0_0 + 5 a^2_0  \big) \, ,   \nonumber \\
 \tau^1_1(s) / m_\pi &= \tfrac{1}{12} \big( 2a_0^0 - 5 a^2_0 \big) \, ,\nonumber \\
 \tau^2_0(s) / m_\pi &= \tfrac{1}{6} \big( 2 a^0_0 +  a^2_0  \big)  \, . 
  \label{tau_gkpy}
\end{align}
In contrast to the \emph{twice-subtracted} equations, the dispersive integrals do not go to zero as $\sqrt{s}\to 2 m_\pi$, and their value in this limit must conspire with the $\tau$ values above to generate the correct threshold behavior. These \emph{minimally-subtracted} equations have two main advantages over the \emph{twice-subtracted} variant. Firstly, the error on the dispersed amplitudes grows less quickly with increasing energy above threshold, which in previous analyses led to lower uncertainties on resonance pole locations and residues~\cite{Garcia-Martin:2011nna}. Secondly, the scattering length values can be determined, rather than needing to be supplied as input. The primary disadvantage of \emph{minimally-subtracted} equations is their sensitivity to the high-energy region of scattering, on which, in this lattice application, we have relatively little constraint. We will return to this later in this section.

\begin{figure*}[!htb]
\includegraphics[width=1.0\textwidth]{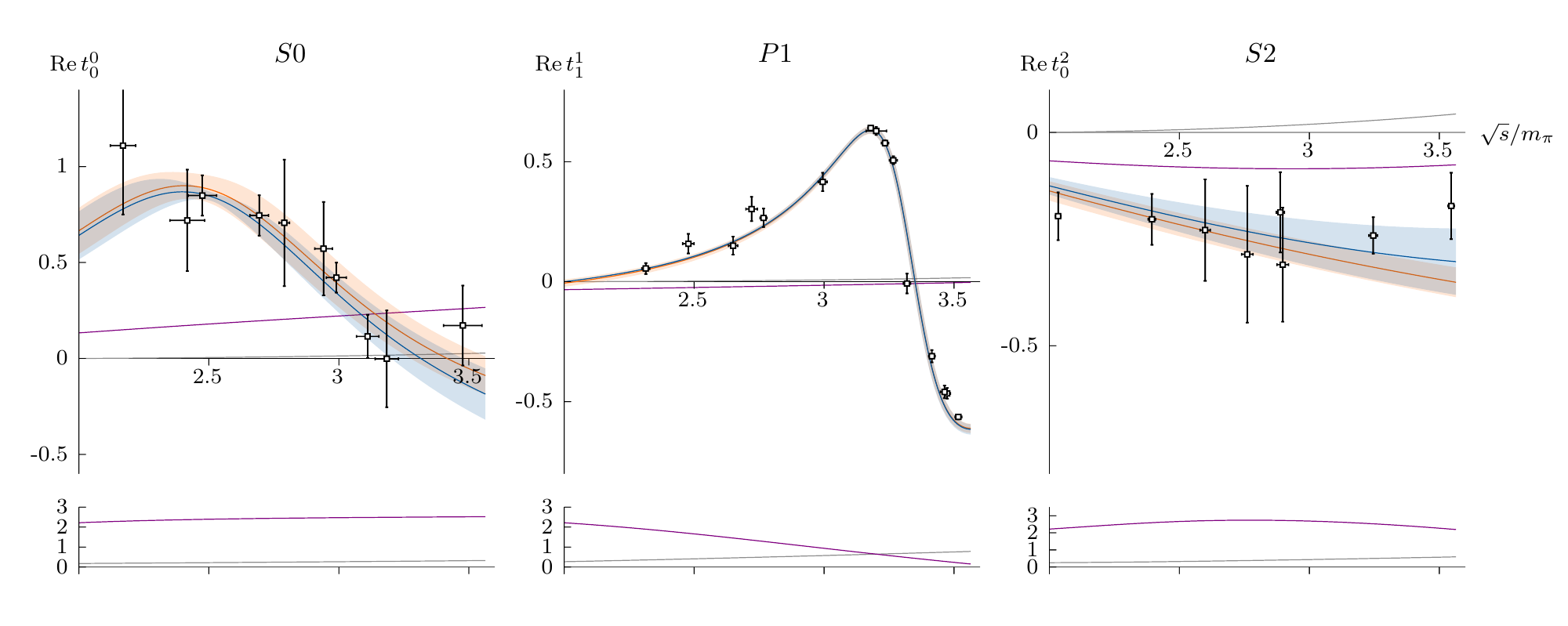}
\caption{ See~\cref{fig:ReggeContrib} for comparison.
Dispersed amplitudes produced by \emph{twice-subtracted} (blue) and \emph{minimally-subtracted} (orange) dispersion relations applied to input amplitudes indicated by the black star in \cref{fig:metric_cuts}. Contributions to each from the high-energy parameterization is shown by the gray and purple curves, respectively.
Bottom panels: Ratio, in absolute value, of $\tilde{f}^I_\ell(s) \big|_\mathrm{Regge}$ and the denominator in the parenthesis of \cref{eq:dsq}, indicating the high-energy contribution in units of the relative uncertainty at each energy, described by the same colors as before.
}\label{fig:GKPY_amplitude}
\end{figure*}

\begin{figure*}[!bth]
\resizebox{\textwidth}{!}{
   \raisebox{-0.5\height}{\includegraphics[width=.6\columnwidth]{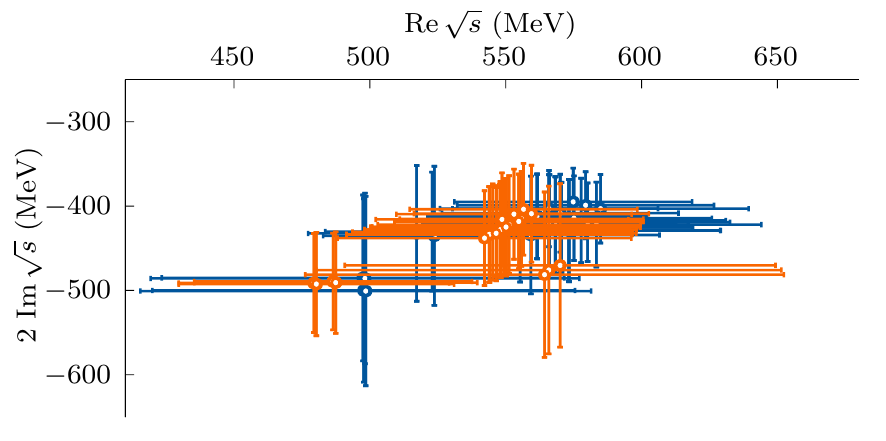}}
    \raisebox{-0.5\height}{\includegraphics[width=.6\columnwidth]{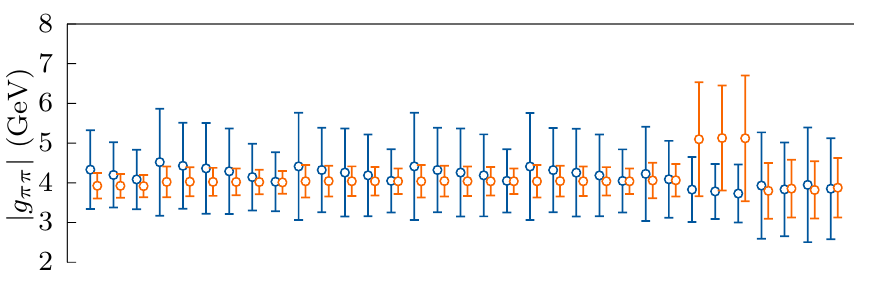}}
   }
\caption{For $m_\pi \sim 239$ MeV. Left: $S0$ pole locations for \emph{twice-subtracted} dispersed amplitudes satisfying the metric cuts described in the text (blue), and for the same amplitude combinations used in \emph{minimally-subtracted} dispersion relations (orange).
Right: Corresponding couplings extracted from the pole residues.
}\label{fig:POLES_GKPY}
\end{figure*}

In the left panel of Figures~\ref{fig:GKPY860}, \ref{fig:GKPY856} we show the $d^2$ and $\tilde{\chi}^2$  metrics for the output of \emph{minimally-subtracted} dispersion relations using the same lattice amplitude parameterizations as shown in Figures~\ref{fig:roy_select860}, \ref{fig:roy_select856}. The black boundary superimposed on the plot is the one selected according to the metric cuts applied to the \emph{twice-subtracted} results, and we see that a large fraction of acceptable solutions of the \emph{minimally-subtracted} dispersion relations also lie in this region. 

In the right panel of \cref{fig:GKPY860} we show a metric designed to show how well the Olsson sum rule~\cite{Olsson:1967vrz}, 
\begin{align}
\label{eq:olsson}
\operatorname{Re} \tilde{T}^{I_t=1}(4m_\pi^2,0) &=\frac{4 m_\pi^2}{\pi} \int_{4 m_\pi^2}^{\infty}  \!ds'\, \frac{\operatorname{Im} T^{I_t=1}(s', 0)}{s'\left(s'-4 m_\pi^2\right)}, 
\end{align}
is satisfied. In this expression, the $t$--channel definite isospin amplitude can be expressed in terms of the $s$--channel definite isospin amplitudes by 
\begin{equation}
    T^{I_t=1}(s, t) = \frac{1}{3}T^{0}(s, t)+\frac{1}{2}T^{1}(s, t)-\frac{5}{6}T^{2}(s, t),
\end{equation} 
and these are expressed in terms of a sum over the partial-wave amplitudes.
The $t$--channel amplitude at threshold is equal to a combination of the $S$--wave scattering lengths,
\begin{equation}
    \operatorname{Re} T^{I_t=1}(4m_\pi^2,0)=\tfrac{32 \pi}{6} (2a^0_0-5a^2_0) \,m_\pi,
\end{equation}
so the quantity plotted, $\left(\frac{\psi}{\Delta \psi}\right)^2$, where
\begin{equation}
\psi = 2 a_0^{0}-5 a_0^{2} - \frac{6 m_\pi}{8 \pi^2} \int_{4 m_\pi^2}^{\infty} \!d s'\, \frac{\operatorname{Im} T^{I_t=1}(s', 0)}{s'\left(s'-4 m_\pi^2\right)} \, ,
\end{equation}
should vanish if this dispersive sum rule is fulfilled.

\vspace{3mm}
Figure~\ref{fig:GKPY_amplitude} shows an example of the dispersed amplitudes coming from \emph{minimally-subtracted} dispersion relations compared to those from \emph{twice-subtracted} dispersion relations for the same set of input lattice amplitude parameterizations, indicated by the black star in \cref{fig:metric_cuts}. 
The amplitudes are observed to be compatible within errors over the entire energy region plotted, but if one examines the contribution to the \emph{minimally-subtracted} amplitudes of the high-energy part of the input, one sees it is much larger than for the \emph{twice-subtracted} variants. Measured as a fraction of the relative uncertainty $\Delta \big[ \tilde{f}^I_\ell(s) - f^I_\ell(s)  \big]$, depicted in the bottom row, it is seen to contribute at a level where one would worry about the correctness of the crudely-scaled Regge parameterization used. As such we do not consider these results to be model-independent consequences of the lattice QCD calculation. Nevertheless, the amplitude agreement with the \emph{twice-subtracted} results inspires us to examine the $\sigma$ pole location, and as shown in \cref{fig:POLES_GKPY}, we see that there is reasonable agreement with the \emph{twice-subtracted} results with, as anticipated, smaller uncertainties on the pole location and the pole residue.

\section{Parameterization functional forms} \label{app:func_forms}

We summarize in this section the functional forms used to describe the input partial-wave amplitudes, with parameters constrained by describing lattice QCD finite-volume energy levels. A similar discussion can be found in Ref.~\cite{Rodas:2023gma} where their application to describe lattice spectra is presented. 

An elastic partial wave can be recast in terms of a real phase-shift as
\begin{equation}
t_{\ell}^{I}(s)=\frac{1}{\rho(s)} \, e^{i \delta^{I}_{\ell}(s)} \sin \delta_{\ell}^{I}(s)=\frac{1}{\rho(s)} \frac{1}{\cot \delta_{\ell}^{I}(s)-i},
\end{equation}
where $\rho(s) = 2 k/\sqrt{s}$ is the two-pion phase-space and ${k = \tfrac{1}{2}\sqrt{ s - 4m_\pi^2}}$ is the scattering momentum. 

At low energies, slow variation of $S$--waves can be described by a low-order expansion in $k^2$, called the \emph{effective range expansion},
\begin{equation}
k^{2\ell+1} \cot \delta_{\ell}^{I} =  F^I_{\ell}(s)\left( \tfrac{1}{a^I_\ell}+\tfrac{1}{2}r^I_\ell k^2+\dots\right),
\end{equation}
where the conventional choice is $F^I_{\ell}(s) = 1$, for which $a_\ell^I$ is interpreted as the \emph{scattering length} and  $r_\ell^I$ as the \emph{effective range}. Additional desired features can be included with other choices of $F^I_{\ell}(s)$, an example being an Adler zero, enforced by using $F^I_\ell(s)= (4m_\pi^2 - s_A)/(s-s_A)$.

Alternatively, we can make use of a \emph{conformal mapping expansion} by defining
\begin{equation}
\Phi^I_{\ell}(s) = \tfrac{2}{\sqrt{s}} k^{2 \ell+1} \cot \delta^I_{\ell}(s),
\end{equation}
which is real analytic between the elastic and inelastic thresholds. One can introduce an effective inelastic threshold, $s_0$, and the opening of the \emph{left-hand-cut} at $s = 0$, by using~\cite{Cherry:2000ut,Pelaez:2016tgi},
\begin{equation}
\omega(s) = \frac{\sqrt{s}-\alpha\sqrt{s_0-s}}{\sqrt{s}+\alpha\sqrt{s_0-s}} \, .
\end{equation}
In this expression, $\alpha$ and $s_0$ are fixed parameters that determine which energy region is mapped into a unit disk of $\omega$. For $S2$ we set $s_0 = 0.09 \,a_t^{-2}$ and $\alpha=0.8$.  For $S0$ we use $\alpha=0.8,\,1$, and consider two values, ${s_0 = 0.032 \,a_t^{-2},\, 0.04 \,a_t^{-2}}$. Expanding the desired function as a polynomial in this variable,
\begin{equation}
\Phi^I_\ell(s)= F^I_{\ell}(s) \sum_{n=0}^N B_{n} \, \omega^{n} \, ,
\end{equation}
we expect convergence with a low-order $N$. Again, one may build in additional properties by using $F^I_\ell(s)$ other than $1$. For example, when a narrow resonance is known to be present, it is convenient to set $F^I_\ell(s) = (s-m^2_R)/m_R^2$. Furthermore, as detailed in Ref.~\cite{Yndurain:2007qm}, spurious singularities appearing below threshold can be removed by adding a function $\gamma^I_\ell(s)$, so that,
\begin{equation}
\Phi^I_\ell(s)= F^I_{\ell}(s) \left(\gamma^I_\ell(s)+\sum_{n=0}^N B_{n} \, \omega^{n} \right)\, .
\end{equation}

Partial-waves dominated by narrow resonances, like $P1$, are typically well-described in the vicinity of the pole by a Breit-Wigner form, 
\begin{equation}
t_{\ell=1}(s)=\frac{1}{\rho(s)} \frac{\sqrt{s}\,  \Gamma(s)}{m_\mathrm{BW}^2 - s - i \sqrt{s}\,  \Gamma(s)} \, ,
\label{eq:bw}
\end{equation}
where the energy-dependent width is given by ${\Gamma(s) = \tfrac{g_\mathrm{BW}^2}{6\pi} \tfrac{k^3}{s}}$. Another, more flexible, parameterization which allows for the presence of resonances, and which generalizes nicely to the case of \emph{coupled-channel} amplitudes, is the $K$-matrix approach~\cite{Aitchison:1972ay},
\begin{equation}
\left( t^I_\ell(s) \right)^{-1} = \left( K^I_\ell(s) \right)^{-1} - i \rho(s) \, ,
\end{equation}
where a common parameterization choice is a sum of poles plus a finite-order polynomial,
\begin{equation}
K^I_\ell(s) = (2k)^{2\ell} \left[ \sum_r \frac{g_r^2}{m_r^2 - s} + \sum_p \gamma_p \, s^p \right] \, .
\end{equation}
This form can be modified to include an Adler zero by taking $K^I_\ell(s) \to (s-s_A)\,  K(s)$. Another choice is to model $\left( K^I_\ell(s) \right)^{-1}$ directly.

The simple phase-space, $\rho(s)$, defined above has an unphysical singularity at $s=0$, but this can be removed by writing a once-subtracted dispersion relation leading to the Chew-Mandelstam function~\cite{Chew:1960iv},
\begin{equation}
I(s)=I(s_M)+\frac{s-s_M}{\pi} \int_{s_{thr}}^{\infty} d s^{\prime} \frac{-\rho\left(s^{\prime}\right)}{(s^{\prime}-s_M)(s^{\prime}-s)} ,
\end{equation}
which has $\mathrm{Im} \, I(s) = - \rho(s)$ above the two-pion threshold, as required by unitarity, and which, when subtracted at threshold, takes the simple form $I(s)=\frac{\rho(s)}{\pi} \log \left[\frac{\rho(s) +1}{\rho(s) - 1}\right]$.

In this work, the input amplitudes constrained by lattice QCD energy levels span a range of forms presented above. The $P1$ amplitudes are fixed to the successful `pole plus constant' $K$-matrix with Chew-Mandelstam phase-space, but many forms are used for the $S0$ and $S2$ amplitudes, leading to a total of over 700 input amplitude combinations being tested within dispersion relations.

\begin{figure*}[!hbt]
\resizebox{\textwidth}{!}{
  \includegraphics{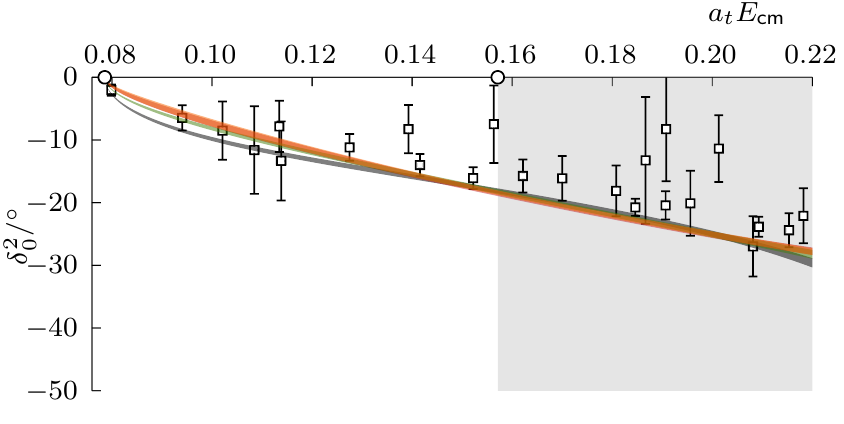} \hspace{.1cm} \includegraphics{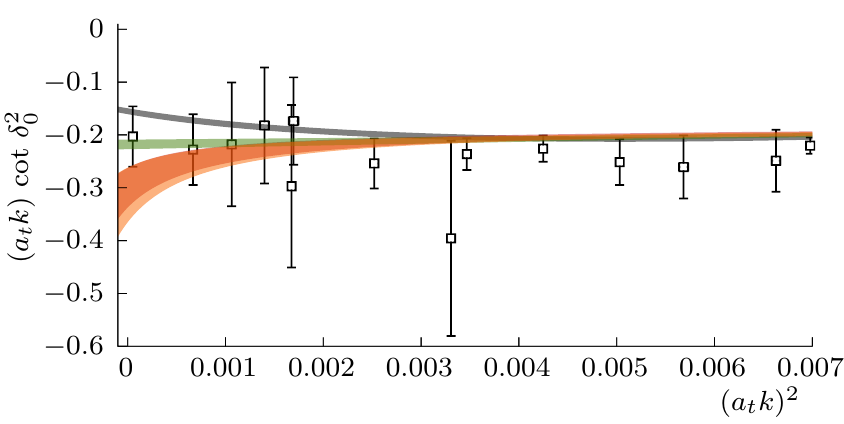} }

  \resizebox{0.5\textwidth}{!}{\includegraphics{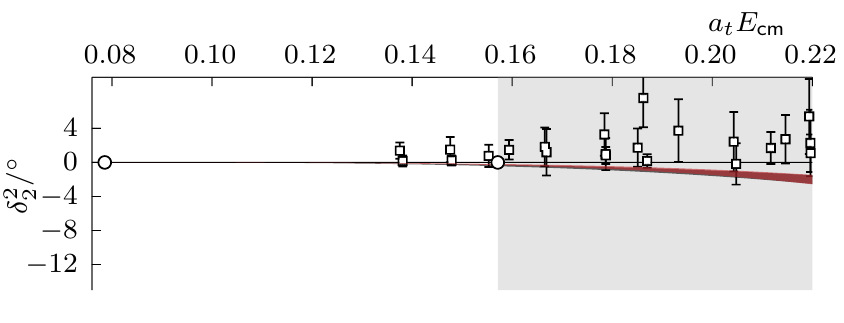} }
\caption{ $I=2$ partial-wave amplitudes for $m_\pi \sim 239$ MeV. For $S$--wave (upper two panels), four example parameterizations are shown: a two-term conformal mapping (black), an effective range expansion with two terms (green), and two choices with an Adler zero fixed at the leading-order $\chi$PT location, a two-term conformal mapping (red), and an effective range expansion with two terms (orange). The $D$--wave phase-shift is shown in the lower panel illustrated with two parameterizations: a scattering length (red), and a conformal mapping with one term (black). Data points with very large uncertainties have been removed from the plots.
}\label{fig:860I2}
\end{figure*}

The $S0$ and $S2$ partial-waves are described by input amplitudes both with and without explicit Adler zeroes. As shown in Ref.~\cite{Garcia-Martin:2011iqs}, dispersive analyses applied to physical $\pi\pi$ scattering data suggest that Adler zeroes may not be located at the tree-level locations. In our input amplitudes, we fix the Adler zeroes either to their tree-level values, or to the extremes of the ``CFD" dispersive range in Ref.~\cite{Garcia-Martin:2011iqs}, extrapolated to our pion masses with $s_A = s_A^\mathrm{phys} \left(m_\pi/m_\pi^\mathrm{phys}\right)^2$. In the case of the $S0$ wave for $m_\pi\sim 283$ MeV, we also consider a couple of parameterizations in which the Adler zero location is allowed to float in the description of the lattice QCD energy levels, and in this case alone we find that this does not degrade significantly the precision of the fit. In other cases, allowing the Adler zero location to float leads to an extremely imprecise estimate of the location.

\section{$m_\pi \sim 239 \, \mathrm{MeV}$ $I=2$ amplitudes}
\label{app:860I2}

    Ref.~\cite{Briceno:2016mjc} and Ref.~\cite{Wilson:2015dqa} reported on extractions of $I=0$ and $I=1$ finite-volume spectra from a $32^3\times 256$ anisotropic Clover lattice with light-quark masses such that $m_\pi \sim 239$ MeV. The corresponding spectra on this lattice with $I=2$ have not previously appeared in the literature, but have been computed for the purposes of this manuscript using the same techniques, range of energies, and similar operators to the ones reported in Ref.~\cite{Rodas:2023gma}. As observed in the previous work, the $D$-wave is very weak and repulsive, and can be described with a single parameter. The $S$-wave shows stronger repulsion, with a phase-shift whose magnitude grows with energy. \cref{fig:860I2} shows discrete values of $S$ and $D$--wave phase-shifts determined from these spectra along with a sample of parameterizations. The parameterizations used here to describe the data correspond to a subsect of those types described above and in Section {\color{jlab_green}III} of~\cite{Rodas:2023gma}. A compilation of the scattering lengths obtained from these fits was presented in Figure~{\color{jlab_green}5} of that work, and their explicit values, in $m^{-1}_\pi$ units, are given as part of~\cref{fig:roy_select860}.

\bibliographystyle{apsrev4-2}
\bibliography{largebiblio.bib,bib.bib}

\end{document}